\documentclass[sigconf, nonacm]{acmart}

\setcopyright{none}
\acmDOI{}

\AtBeginDocument{%
  \providecommand\BibTeX{{%
    \normalfont B\kern-0.5em{\scshape i\kern-0.25em b}\kern-0.8em\TeX}}}

\usepackage{threeparttable}
\usepackage{balance}

\usepackage{url}
\usepackage{breakurl}

\usepackage{multirow}
\usepackage{enumitem}

\usepackage[nounderscore]{syntax}

\usepackage{algorithmic,float}
\usepackage[noend, ruled, linesnumbered]{algorithm2e}

\usepackage{booktabs}
\usepackage{utfsym}
\usepackage{microtype}
\usepackage{multicol}
\usepackage{tcolorbox}

\usepackage{listings}
\usepackage{tikz}
\usepackage{amsmath}
\usepackage{xcolor}
\usepackage{soul}
\usepackage{makecell}
\usepackage{multicol}
\usepackage{xspace}
\usepackage{balance}
\usepackage{pdfpages}
\usetikzlibrary{positioning,arrows, decorations.markings,arrows.meta,fit,calc,shapes.geometric,bending,positioning}

\lstdefinelanguage{cypher}{
  morekeywords = {WITH, ORDER, BY, CALL, UNWIND, RETURN, AS, FOREACH, IN, MERGE, EXISTS, MATCH, CREATE, none, WHERE, ON, SET, DELETE, SELECT, FROM, CASE, WHEN, THEN, END, UNION, OR, AND, IS, NOT, REMOVE, DETACH, OPTIONAL, ELSE, COUNT, SHOW, FUNCTIONS, USE},
  morecomment=[l]{//},
}
\newcommand{\cypher}{\lstinline[language=cypher]}

\usepackage{listings}
\definecolor{mygreen}{rgb}{0,0.6,0}
\definecolor{mygray}{rgb}{0.98,0.98,0.98}
\definecolor{mymauve}{rgb}{0.58,0,0.82}
\definecolor{myblue}{rgb}{0.2,0.42,0.62}
\definecolor{myred}{rgb}{0.75,0,0}
\usepackage{soul} 
\sethlcolor{pink}
\newcommand{\highlightedFOREACH}{\textcolor{myblue}{\textbf{FOREACH}}}
\newcommand{\highlightedIN}{\textcolor{myblue}{\textbf{IN}}}
\newcommand{\highlightedSET}{\textcolor{myblue}{\textbf{SET}}}
\newcommand{\highlightedCREATE}{\textcolor{myblue}{\textbf{CREATE}}}
\newcommand{\highlightedDELETE}{\textcolor{myblue}{\textbf{DELETE}}}

\newcommand{\highlightedRETURN}{\textcolor{myblue}{\textbf{RETURN}}}
\newcommand{\highlightedAS}{\textcolor{myblue}{\textbf{AS}}}
\newcommand{\highlightedEXISTS}{\textcolor{myblue}{\textbf{EXISTS}}}

\usepackage{tcolorbox}

\newcommand{\tikzcheckmark}{%
    \begin{tikzpicture}[scale=0.2]
        \draw[line width=2pt] (-0.2,0.2) -- (0.3,-0.3) -- (1.3,0.7);
    \end{tikzpicture}%
}

\newcommand{\bugmark}{
  \hspace{-3px}\raisebox{-2px}{\begin{tikzpicture}[scale=0.11]
    \fill[red] (0,0) circle (1cm);
    \draw[black, thin] (0,-1) -- (0,1);
    \fill[black] (0,0.8) circle (0.6cm);
    \fill[black] (-0.5,0.6) circle (0.2cm);
    \fill[black] (0.5,0.6) circle (0.2cm);
    \fill[black] (-0.65,0) circle (0.15cm);
    \fill[black] (0.65,0) circle (0.15cm);
    \fill[black] (-0.45,-0.55) circle (0.15cm);
    \fill[black] (0.45,-0.55) circle (0.15cm);
    \draw[black, thick] (-0.2,1.2) .. controls (-0.3,1.5) and (-0.6,1.8) .. (-0.8,1.6);
    \draw[black, thick] (0.2,1.2) .. controls (0.3,1.5) and (0.6,1.8) .. (0.8,1.6);
    \draw[black, thick] (-0.8,0.6) -- (-1.3,1.1);
    \draw[black, thick] (-1,0) -- (-1.5,0); 
    \draw[black, thick] (-0.8,-0.6) -- (-1.3,-1.1);
    \draw[black, thick] (0.8,0.6) -- (1.3,1.1);
    \draw[black, thick] (1,0) -- (1.5,0); 
    \draw[black, thick] (0.8,-0.6) -- (1.3,-1.1);
  \end{tikzpicture}}
}

\newcommand{\bugsfound}{127\xspace}
\newcommand{\bugsfixed}{84\xspace}
\newcommand{\bugsconfirmed}{113\xspace}
\newcommand{\bugslogic}{33\xspace}

\newcommand{\sys}{\mbox{\textsc{Dinkel}}\xspace}

\SetCommentSty{mycommfont}

\usepackage{pifont}

\newcommand{\filledcircle}{\tikz\draw[black,fill=black] (0,0) circle (.75ex);} 
\newcommand{\emptycircle}{\tikz\draw[black] (0,0) circle (.75ex);} 

\newcommand{\midspace}{0.5ex}
\newcommand{\bigspace}{1ex}

\begin{document}

\title[Dinkel: State-Aware and Granular Framework for Validating Graph Databases]{Dinkel: State-Aware and Granular Framework for \\
Validating Graph Databases
}

%
\author{Celine Wüst}
\authornote{Both authors contributed equally to this research.}
\affiliation{%
  \institution{ETH Z\"urich}
  \country{Switzerland}
}

\author{Zu-Ming Jiang}
\authornotemark[1]
\affiliation{%
  \institution{ETH Z\"urich}
  \country{Switzerland}
}
\author{Zhendong Su}
\affiliation{%
  \institution{ETH Z\"urich}
  \country{Switzerland}
}


\begin{abstract}
Graph database management systems (GDBMSs) have been powering many data-driven applications. 
To ensure GDBMS reliability, 
several testing approaches have been proposed.
However, they all suffer from two key limitations: (1) insufficient support for generating complex and valid queries to exercise deep GDBMS code, and (2) lack of general oracles to validate the execution correctness of arbitrary queries.

In this paper, we propose a novel and practical approach \sys, to thoroughly test GDBMSs. Our approach consists of two core techniques.
First, to generate complex and valid queries, we model two kinds of graph state, \emph{query context} and \emph{graph schema}, to describe the Cypher variables and the manipulated graph labels and properties. We generate queries clause-by-clause, and modify the graph states on the fly to ensure each clause references the correct state information.
Second, to generally validate query results, we introduce two fine-grained query transformations: clause-level and expression-level transformations.
These transformations can operate on arbitrary queries while preserving their semantics. \sys validates GDBMSs by checking whether the transformed query produces the same results as the original.
We evaluated \sys on three well-known GDBMSs. In total, we found \bugsfound bugs, among which \bugsconfirmed were confirmed, \bugsfixed were fixed, and \bugslogic were logic bugs.
Compared to existing approaches, \sys can cover over 70\% more code and find substantially more bugs within a 48-hour testing campaign.
We expect \sys's powerful bug detection to lay a practical foundation for GDBMS testing. 


\end{abstract}

\ccsdesc[500]{Information systems~Query languages}
\ccsdesc[500]{Security and privacy~Database and storage security}



\maketitle

\section{Introduction}
\label{sec_intro}

Graph database management systems (GDBMSs) are crucial for modern interconnected, data-driven computer software~\cite{not-lost-case-study, novartis-case-study, generative-ai-neo4j, nbc-case-study, yin2023asplos, chen2023asplos, xu2019asplos}.
Notably, 75\% of the Fortune 100 and all of North America's top 20 banks have adopted the currently most popular GDBMS, 
Neo4j~\cite{db-ranking-dbms, who-uses-neo4j}.
While rapidly evolving and becoming more complex (\emph{e.g.}, Neo4j has 1.2M LOC), GDBMSs inevitably suffer from bugs introduced during their development.
These bugs are critical as they can corrupt the GDBMSs and lead to serious consequences. One of the most critical types of bugs are logic bugs, which can silently cause GDBMSs to produce incorrect query results.

To improve GDBMS reliability, 
testing approaches~\cite{ziyue2023issta, matteo2023issta, jiang2023icse, zhuang2023vldb} have been proposed to find bugs by generating queries in Cypher, the most widely-adopted graph query language~\cite{opencypher}. As large-scale, complex systems, GDBMSs are difficult to test effectively due to two key challenges: \emph{query generation} and \emph{test-oracle construction}.
To cover a broad spectrum of functionalities and deep system logic, approaches need to generate complex queries involving various Cypher language features and complicated data dependencies. To identify bugs, especially logic bugs that cause GDBMSs to silently produce incorrect results for queries, testing approaches require test oracles that can be generally applied to validate arbitrary queries. 

However, both of these challenges remain unsolved. All existing approaches focus on proposing new test oracles, but struggle to systematically generate complex and valid queries, leaving much GDBMS logic and code inadequately tested or completely unexercised. Although existing approaches focus on test oracles to identify logic bugs, none of the proposed oracles can generally be applied to validate arbitrary queries. As a result, many logic bugs in GDBMSs get missed, even though they can be triggered by existing queries. The following discusses the details of these two challenges.

\vspace{\bigspace}
\noindent
\underline{\textbf{Challenge 1:} \textit{Generating Complex Queries.}}
\,
In DBMSs, many bugs can only be triggered by complex queries~\cite{zuming2023sec, zuming2024osdi, zuming2023osdi}. A complex Cypher query can involve various clause combinations and complicated data dependencies.
Generating such complex queries is challenging because the clauses invoked by Cypher queries can change the graph states of the manipulated graph database during their execution. Such state changes are visible in subsequent clauses. Moreover, the data and variables used by Cypher queries have different scopes, depending on the clauses they are involved in.
Due to the lack of systematic modeling of these Cypher language features, existing approaches cannot effectively generate complex queries, and thus, many critical functionalities and deep logic of GDBMSs are not exercised by the testing campaigns of these approaches.
Therefore, many deep bugs (\emph{e.g.}, the bugs shown in Figure~\ref{fig:assertion_failure_redisgraph} and Figure~\ref{fig_neo4j_bug_intro}) in GDBMSs are missed by existing approaches.

Modeling Cypher language features is not trivial.
Cypher is a declarative language that allows expressive data querying without describing specific control flows, which enables GDBMSs to flexibly select efficient strategies to execute queries.
However, different from typical declarative query languages (\emph{e.g.}, SQL), Cypher queries can make changes to the graph database state while executing Cypher clauses of the queries, whose effects are visible in subsequent clauses~\cite{opencypher-reference}. For example, a graph node created in an outer \cypher{CREATE} clause can be referenced by subsequent \cypher{DELETE} clauses. 
Such characteristics make it challenging to effectively test GDBMSs. On the one hand, neglecting these characteristics tends to make the generated queries simpler, resulting in the GDBMSs not having to handle complicated data dependencies and therefore not allowing the testing campaign to reach the deep logic of GDBMSs.
On the other hand, failing to correctly handle such characteristics can make the generated Cypher queries invalid, resulting in many queries being discarded by GDBMSs in the early stages (\emph{e.g.}, query parsing). For example, the generated queries may become invalid when referencing nonexistent or out-of-scope items.

\begin{figure}[t]
    \begin{lstlisting}[language=cypher, 
    ]
MERGE ((|\color{mygreen}{x}|))<-[(|\color{violet}{:A}|)]-((|\color{mygreen}{x}|))<-[(|\color{violet}{:A}|)]-((|\color{mygreen}{x}|))<-[(|\color{violet}{y:A}|)]->((|\color{mygreen}{x}|))
DELETE (|\color{violet}{y}|)
CREATE ((|\color{mygreen}{x}|))<-[(|\color{violet}{:B}|)]-()
DELETE (|\color{violet}{y}|);
    \end{lstlisting}
    \caption{Assertion failure found by \sys in RedisGraph.}
    \label{fig:assertion_failure_redisgraph}
\end{figure}

\vspace{\bigspace}
\noindent
\underline{\textbf{Challenge 2:} \textit{General Test Oracle.}}
\,
Existing approaches use query transformations to validate the correctness of query results~\cite{matteo2023issta, jiang2023icse, zhuang2023vldb, mang2024icse}. Given a generated Cypher query, they transform it into another query and check whether the results of the transformed query and the original query follow specified relationships.
However, these approaches cannot be generally applied to arbitrary queries because they require their generated queries to follow specific query patterns so that their transformations can be applied.
For example, GraphGenie~\cite{jiang2023icse} requires the predicate in \cypher|MATCH| clauses to satisfy the preconditions of their transformation rules (\emph{e.g.}, one rule requires the predicates to involve a cyclic graph). 
Moreover, GraphGenie can transform and validate only queries that contain a single \cypher{MATCH} clause followed by a \cypher{RETURN}. 
GRev~\cite{mang2024icse} integrates an abstract syntax graph (ASG) for interpreting the predicate in \cypher|MATCH| statements and can thus transform the predicate in more diverse ways and with fewer restrictions.
However, the predicate modeling in GRev can only be applied to \cypher|MATCH| statements, while queries involving other common clauses like \cypher|MERGE|, \cypher|UNWIND|, and \cypher|FOREACH| remain unsupported and thus untested.

Such inapplicabilities arise from the coarse-grained transformation used by existing approaches. Specifically, their transformations are at the query level.
When transforming a query to another related query, these approaches need to take into account the semantics of the entire query and choose transformation rules according to these parsed semantics. They cannot handle semantics that do not fit their transformation rules (\emph{e.g.}, both GraphGenie and GRev do not support \cypher{FOREACH} clauses).
Such coarse-grained transformations make existing approaches inapplicable to various queries that do not match specific query patterns. Therefore, many logic bugs are missed.
For example, existing approaches miss the bug shown in Figure~\ref{fig_neo4j_bug_intro} because the original query is too complex and does not follow any specified patterns.


\begin{figure}[t]
\noindent
\begin{minipage}[t]{0.47\linewidth}
\begin{lstlisting}[
    language=cypher, 
    basicstyle=\footnotesize\tt,
]                              
// (|\textbf{Original: 2 rows}|) (|\bugmark|)
(|\hl{\highlightedRETURN{} 0 \highlightedAS{} n1}|)
UNION CALL {
  FOREACH (n2 IN [] | 
    FOREACH (n3 IN [] | 
      MERGE (:A {u:0})))
} MATCH (y) 
RETURN y AS n1
UNION CREATE () 
RETURN 0 AS n1
  
  
  
\end{lstlisting}
\end{minipage}%
\hfill
\begin{minipage}[t]{0.47\linewidth}
    \begin{lstlisting}[
    language=cypher, 
    basicstyle=\footnotesize\tt,
]          
// (|\textbf{Transformed: {1 row}}|) (|\textcolor{mygreen}{\tikzcheckmark}|)
(|\hl{\highlightedFOREACH{} (n0 \highlightedIN{} [] \textbar{} }|)
  (|\hl{\highlightedFOREACH{} (x \highlightedIN{} [] \textbar{} }|) 
    (|\hl{\highlightedSET{} x = \{\}))}|)
(|\hl{\highlightedRETURN{} 0 \highlightedAS{} n1}|)
UNION CALL {
  FOREACH (n2 IN [] | 
    FOREACH (n3 IN [] | 
      MERGE (:A {u:0})))
} MATCH (y) 
RETURN y AS n1
UNION CREATE ()
RETURN 0 AS n1
\end{lstlisting}
\end{minipage}
\caption{Queries exposing a logic bug in Neo4j.}
\label{fig_neo4j_bug_intro}
\end{figure}

In this paper, we propose a novel and practical approach, \sys, which is designed to lay a practical foundation for testing GDBMSs. \sys integrates two technical solutions to address the two challenges discussed above.

\vspace{\bigspace}
\noindent
\underline{\textbf{Solution 1:} \textit{State-Aware Query Generation.}}
\,
\sys models graph states and state changes caused by Cypher queries. It abstracts the states into two categories, \emph{query context} and \emph{graph schemas}. Query context contains information related to temporary variables declared in the queries (\emph{e.g.}, the type and scope of each variable), while graph schemas maintain the current graph information, including the graph labels and properties. 
As query context and graph schemas may change at different Cypher clauses, we must update these abstractions while generating Cypher queries. To this end, we propose \emph{state-aware query generation}. Instead of determining query skeletons for later expressions complementarily, our approach incrementally constructs clauses for the generated queries. When constructing a new Cypher clause, our approach references only the accessible elements according to the current query context and graph schema. 
After the clause is completed, the approach updates the corresponding state information. In this on-the-fly way, our approach can accurately maintain the dynamically evolving graph state and thus generate queries involving complicated data dependencies and state changes.

\vspace{\bigspace}
\noindent
\underline{\textbf{Solution 2:} \textit{Fine-Grained Query Transformation.}}
\,
Our idea to validate Cypher queries is inspired by EET~\cite{zuming2024osdi}, which is designed for SQL queries in relational DBMSs (RDBMSs). EET proposes to validate SQL queries by transforming them at the level of expressions.
In this paper, we further generalize this method to \emph{fine-grained query transformation}, without limiting the transformation units to just expressions. 
Specifically, instead of analyzing the semantics of a whole query, our approach transforms queries by their fine-grained units. \
In the context of Cypher queries, the fine-grained units can not only be expressions, but also clauses, which can flexibly change graph states.
By operating on these low-level units, we focus on the fine-grained semantics of queries, without the need to analyze the overall query-level semantics.
For example, we can transform the first \cypher|RETURN| clause of the original query in Figure~\ref{fig_neo4j_bug_intro} by inserting an additional \cypher|FOREACH| clause, which is dead code and has no effects because the array to be iterated over is empty.
Therefore, the clauses before and after transformations should be semantically equivalent, and the transformed query should produce the same results as the original one. However, their results differ, indicating that a logic bug was triggered.
In this process, \sys only needs to ensure the semantic equivalence between clauses, without understanding the semantics of the whole query.

\begin{table*}[t]
\caption{Grammar of Cypher query language}
\label{tbl_cypher_grammar}
\noindent{\small
\begin{center}
\begin{tabular}{@{}lll@{}}
\toprule
\textit{query} & ::= & \textit{clause} [\textit{query}] \\
\textit{clause} & ::= & \textit{reading\_clause} | \textit{writing\_clause} | \textit{reading/writing\_clause} | \textit{projecting\_clause} | ...\\
\textit{reading\_clause} & ::= & [\lstinline|'OPTIONAL'|] \lstinline|'MATCH'|  \textit{pattern} [\lstinline|'WHERE'| \textit{expression}] \\ 
\textit{pattern} & ::= & \textit{node} [\textit{relationship pattern}] \\
\textit{node} & ::= & \lstinline|'('| \textit{label}* \textit{properties}? \lstinline|')'|\\
\textit{relationship} & ::= & \lstinline|'<-['| \textit{label} \textit{properties}? \lstinline|']-'| | \lstinline|'-['| \textit{label} \textit{properties}? \lstinline|']->'|\\
\textit{label} & ::= & \lstinline|':'| \textit{identifier} \\
\textit{properties} & ::= & \lstinline|'{'| \textit{identifier} : \textit{expression} [\lstinline|,| \textit{identifier} : \textit{expression}] \lstinline|'}'| \\
\textit{expression} & ::= & \textit{identifier} | \textit{constant} | \textit{operation} | \textit{function} | ... \\
\textit{writing\_clause} & ::= & \textit{create\_clause} | \textit{delete\_clause} | \textit{set\_clause} | \textit{remove\_clause} | \textit{foreach\_clause} | ...\\
\textit{reading/writing\_clause} & ::= & \textit{merge\_clause} | \textit{call\_clause} | ... \\
\textit{projecting\_clause} & ::= & \textit{return\_clause} | \textit{with\_clause} | \textit{unwind\_clause} | ...\\
\bottomrule
\end{tabular}
\end{center}}
\end{table*}

Regarding the transformation for clauses, \sys manipulates target clauses to affect the graph state in ways that do not interfere with the semantics of the following clauses (\emph{e.g.}, creating and immediately deleting a node), or introduce dead code (\emph{e.g.}, a \cypher|MATCH| clause matching a node that does not exist) before target clauses.
Regarding the transformation for expressions, given an arbitrary expression within clauses, \sys can construct a semantically equivalent expression by leveraging logical (\emph{e.g.}, De Morgan's laws) and algebraic equivalences (\emph{e.g.}, $x \equiv x + 0$). \sys also uses features specific to the Cypher language for transforming expressions (\emph{e.g.}, \cypher|x| $\equiv$ \cypher|CASE WHEN true THEN x END|).

Ultimately, we implemented \sys as a fully automatic GDBMS testing framework.
\sys will be open-sourced once the paper is ready to be public.
We have evaluated our tool on three popular GDBMSs (\emph{i.e.}, Neo4j~\cite{neo4j-repo}, RedisGraph~\cite{redisgraph-repo}, and Memgraph~\cite{memgraph-repo}).
In the evaluation, \sys efficiently generated complex Cypher queries and kept a high validity rate (89.02\%).
Using these queries,
\sys found \bugsfound unique, previously unknown bugs, among which \bugslogic are logic bugs.
So far, \bugsconfirmed of these bugs have been confirmed, and \bugsfixed fixed. 
Many bugs are long-latent and missed by all existing approaches during their extensive evaluation.
Compared to existing approaches, \sys can cover over 70\% more code and find more bugs within the 48-hour testing campaign.
These results demonstrate that \sys significantly outperforms existing approaches in bug detection for GDBMSs.

In summary, we make the following contributions:

\begin{itemize}[leftmargin=*,] 

\item \textbf{\textit{Novel approach:}}
We tackle two fundamental problems of GDBMS testing via two novel solutions: (1) \emph{state-aware query generation}, which abstracts the Cypher state information to facilitate query generation;
and (2) \emph{fine-grained query transformation}, which can be generally applied to validate arbitrary Cypher queries by manipulating their clauses and expressions.

\item \textbf{\textit{Practical realization:}} We realize a fully automatic testing framework, \sys, to find bugs in GDBMSs by generating and validating complex Cypher queries.

\item \textbf{\textit{Promising results:}}
We evaluated \sys on 
three popular open-source GDBMSs, namely 
Neo4j, RedisGraph and Memgraph. \sys found \bugsfound unique and new bugs, among which \bugslogic are logic bugs.
So far, \bugsconfirmed unique bugs have been confirmed, and \bugsfixed fixed. These results demonstrate the effectiveness of \sys.

\end{itemize}

\section{Background}

\noindent
\textbf{Graph Database Management Systems}.
GDBMSs utilize graph models to represent data. The most widely used graph model is the \emph{labeled property graph model}~\cite{jiang2023icse}, which stores interconnected data using nodes connected via relationships (\emph{i.e.}, directed edges between nodes)~\cite{opencypher}. 
Nodes and relationships, generally referred to as \emph{graph entities}, can be associated with labels and properties.
Labels are used to group and classify elements, whereas properties are made up of key-value pairs, providing attribute information.

\vspace{\midspace}
\noindent
\textbf{Cypher Language.}
Cypher is the most widely adopted query language for property graph databases~\cite{opencypher} and was proposed by Neo4j. It is designed as a declarative query language, which allows users to specify the required data without realizing the detailed procedures. 
The general way to specify data in Cypher is to concretize the \emph{graph patterns}, which follow the format \cypher{(n)-[r]->(m)} and can be used to reference graph entities satisfying specified conditions. 
For example, the query in Figure~\ref{fig:assertion_failure_redisgraph} 
concretizes a graph pattern following the \cypher|CREATE| clause to specify the graph entities to be created.

Unlike the query language SQL, Cypher does not differentiate between data declaration (DDL), manipulation (DML), or query (DQL) languages.
Instead, a Cypher query can create, read, and modify data in a single query, thereby allowing queries of procedural nature and non-trivial state manipulation. Generally, Cypher queries access or operate on graphs via clauses, and each query can contain multiple clauses.


Table~\ref{tbl_cypher_grammar} shows the context-free grammar of the Cypher query language. A Cypher query consists of a sequence of clauses. Each clause can be either a reading clause, writing clause, reading/writing clause, projecting clause, or others (\emph{e.g.,} system configuration clause)~\cite{opencypher-reference}. Reading clauses (\emph{i.e.}, \cypher|MATCH|) query the GDBMS to extract information without modifying the graph entities. Writing clauses (\emph{e.g.}, \cypher|CREATE|, \cypher|DELETE|) modify the data stored in GDBMSs by changing the nodes or relationships in the graph. Reading/writing clauses (\emph{e.g.}, \cypher|MERGE|) can both read and write data in the graph. Projecting clauses (\emph{e.g.}, \cypher|WITH|, \cypher|RETURN|) define expressions to be referenced in the subsequent clauses or the result set. When executing a query with multiple clauses, GDBMSs process these clauses sequentially, and the graph state may change after each clause is processed.

\begin{figure}[t]
    \begin{center}
        \colorlet{ggray}{black!10}

\sffamily

\begin{tikzpicture}
    \tikzset{node/.style={black, line width=0.5mm, circle, draw, minimum size=.7cm, fill=ggray}}
    \tikzset{edge/.style={-Latex, black, line width=0.3mm}}

    \node[node] (a0) {\large x};
    \draw[edge] (a0) to[out=105, in=75, loop] node[xshift=-.1cm, yshift=-.2cm, left] {\lstinline{y}:A} (a0);
    \draw[edge] (a0) to[out=225, in=195,loop] node[left]  {:A} (a0);
    \draw[edge] (a0) to[out=345, in=315,loop] node[right] {:A} (a0);
    
    \node[node] (a1) [right=1.7cm of a0] {\large x};
    \draw[edge] (a1) to[out=105, in=75  ,loop] node[xshift=-.1cm, yshift=-.2cm, left] {:A} (a1);
    \draw[edge] (a1) to[out=-75, in=-105,loop] node[xshift=.1cm, yshift=.2cm, right] {:A} (a1);
    
    \draw[-{Triangle[length=.3cm,angle'=95]}, black, line width=.3cm] ([xshift=.9cm]a0.east) -- ([xshift=-.3cm]a1.west);

    \node[node, yshift=-.55cm] (a2) [right=1.5cm of a1] {\large x};
    \node[node] (a3) [above=0.5cm of a2] {};
    \draw[edge] (a2) to[out=-15, in=-45,loop] node[right] {:A} (a2);
    \draw[edge] (a2) to[out=-135, in=-165,loop] node[left] {:A} (a2);
    \draw[edge] (a3.south) -- node[yshift=.1cm, right] {:B} (a2.north);
    
    \draw[-{Triangle[length=.3cm,angle'=95]}, black, line width=.3cm] ([xshift=.3cm]a1.east) -- ([xshift=.8cm]a1.east);

    \node (b0) [above=.8cm of a0]             {\small \textbf{Line 1}};
    \node (b1) [above=.8cm of a1]             {\small \textbf{Line 2}};
    \node[yshift=1cm] (b2) [above=.35cm of a2] {\small \textbf{Line 3-4}};
\end{tikzpicture}
    \end{center}
    \vspace{-5pt}
    \caption{State changes of the query in Figure~\ref{fig:assertion_failure_redisgraph}.}
    \label{fig:statechange} 
\end{figure}

\vspace{\midspace}
\noindent
\textbf{Illustrative Example.}
Figure~\ref{fig:assertion_failure_redisgraph} shows a query with multiple clauses. This query triggers an assertion failure in RedisGraph. 
Figure~\ref{fig:statechange} shows the corresponding graph state changes during execution of each clause.
The first clause, \cypher{MERGE}, will create graph entities following the specified graph pattern if no graph entity matches the pattern. Therefore, a node \cypher{x} and three relationships with labels \cypher{A} are created. Each relationship is from node \cypher{x} to node \cypher{x}. The subsequent \cypher{DELETE} checks whether the graph entity \cypher|y| exists and deletes the existing ones, ultimately deleting one relationship.
The second \cypher{CREATE} clause creates a new node connected to node \cypher{x} with a new relationship \cypher{B}. The last clause, \cypher{DELETE}, tries to delete relationship \cypher{y}, but as \cypher{y} has already been deleted, this \cypher{DELETE} should do nothing. However, RedisGraph recognized \cypher{y} as non-deleted and performed a deletion on a nonexistent relationship, which triggered an assertion failure.
Such a mistake is caused by the entity ID reuse mechanisms of RedisGraph, which reassigns the ID of the deleted relationship \cypher{y} to the new relationship created by the \cypher{CREATE} clause. When referencing \cypher{y} in the last \cypher{DELETE} clause, RedisGraph mistakenly considers \cypher{y} to still exist because its entity ID is being used.

Cypher queries with multiple clauses changing graph states are commonly used in the real world due to the complex nature of data relationships within graph databases.
However, it is extremely challenging to automatically generate and validate such queries, because
(1) the generation needs to be aware of the graph state changes caused by each clause (\emph{e.g.}, \cypher|MERGE| and \cypher|DELETE|) and properly reference the intermediate data (\emph{e.g.}, variables \cypher|x| and \cypher|y|); and (2) the validation needs to construct general test oracles in the case of the complicated semantics the query can contain.

\section{Approach}

\begin{figure}
\centering\includegraphics[width=0.8\linewidth]{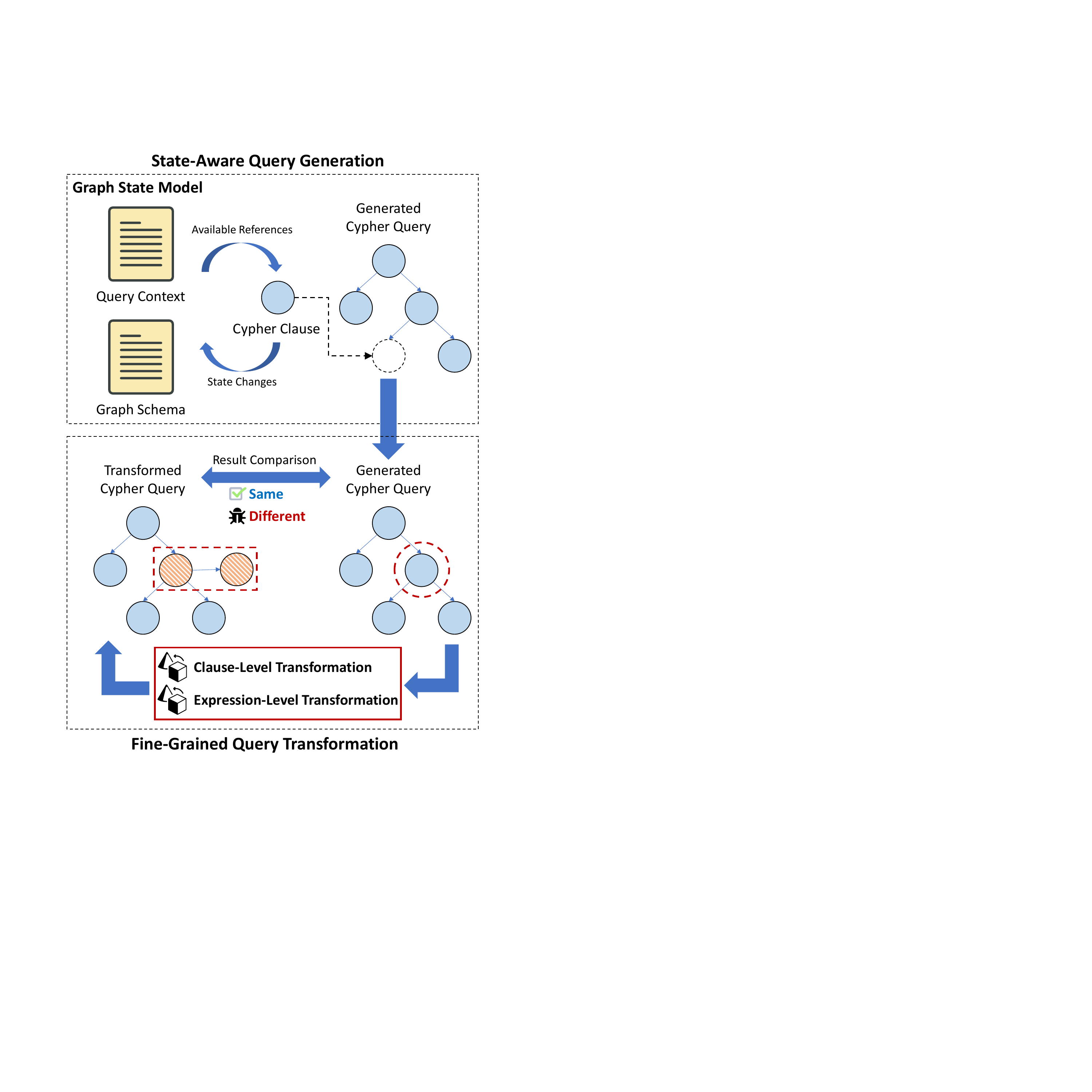}
        \vspace{-5pt}
	\caption{Approach overview.} 
	\label{fig_approach_overview}
        \vspace{-5pt}
\end{figure}

We propose \sys, a GDBMS testing framework integrating two techniques: state-aware query generation and fine-grained query transformation. Figure~\ref{fig_approach_overview} shows its overview.

\subsection{State-Aware Query Generation}

\sys introduces two abstractions, \emph{query context} and \emph{graph schema}, to precisely model the graph states maintained by GDBMSs.
Based on these abstractions, \sys incrementally constructs Cypher queries clause by clause. Each time a new clause is being constructed, our approach checks the current query context and graph schema, builds the clause with the available data references, and updates the state information accordingly.
In this way, \sys continuously constructs clauses for generating complex queries while accurately maintaining the graph state information to introduce complicated data dependencies.

\subsubsection{Graph State Modeling.}

Cypher query execution can be affected by two kinds of graph states:

\vspace{\midspace}
\noindent
\textbf{Query Context.} 
One kind of graph state are the temporary variables declared in the query. These variables have specific types and scopes, affecting only the query state where they are declared. They can either be concrete values with specific types (\emph{e.g.}, integer 0) or aliased to specific nodes or relationships (\emph{e.g.}, node \cypher{x} and relationship \cypher{y} in Figure~\ref{fig:assertion_failure_redisgraph}). Such variables can be referenced only after they are defined, within their scope.
We refer to these variables, their types, and their scope information as \emph{query context}.
Query context may change at different Cypher clauses. Specifically, query context includes new information when a new variable is declared by a clause, and excludes outdated information when the clause is out of the scope of existing variables. 

\vspace{\midspace}
\noindent
\textbf{Graph Schema.}
Another graph state is the schema of the stored graph data. It includes graph labels and properties (\emph{e.g.}, label \cypher{A} and \cypher{B} for relationships in Figure~\ref{fig:assertion_failure_redisgraph}). We refer to such state information as \emph{graph schema}. Graph schemas can be changed by specific clauses. For example, \cypher{CREATE} clauses can create nodes or relationships with new labels and properties, while \cypher{REMOVE} clauses can remove existing labels or properties.
Different from query context, whose effects are limited by their scope, the operations (\emph{e.g.}, \cypher{CREATE} clause that declares new labels) on graph schema affect the database permanently. In addition, graph schema can be referenced even though the labels and properties are nonexistent. 
This design improves query flexibility as users can write valid queries without concerning themselves with the current graph schema. 

\subsubsection{Query Generation}
\label{sec_state_aware_query_generation}
Our generation is based on a key insight that clauses are the basic units for operating on graphs, and both query context and graph schema are updated only when clauses are invoked or exit specific clause context. Based on this insight, our approach track the state changes inside Cypher queries by analyzing the possible impact caused by specific clauses in queries.

\begin{algorithm}[t]
    \SetKwData{P}{P}
    \SetKwData{L}{L$_{max}$}
    \SetKwData{any}{ANY}
    \SetKwData{ClauseType}{CLAUSE\_TYPE}

    \SetKwInOut{Input}{input}\SetKwInOut{Output}{Output}
    \SetKwProg{Fn}{Function}{:}{}
    \SetKwFunction{GenQuery}{GenQuery}
    \SetKwFunction{EmptyQuery}{EmptyQuery}
    \SetKwFunction{GenClause}{GenClause}
    \SetKwFunction{GenExpression}{GenExpression}
    \SetKwFunction{append}{AppendClause}
    \SetKwFunction{RandClauseType}{RandClauseType}
    \SetKwFunction{InitializeClause}{RandInitClause}
    \SetKwFunction{RandGenExpr}{RandGenExpr}
    \SetKwFunction{CleanQC}{CleanQC}
    \SetKwFunction{length}{Length}
    \SetKwFunction{rand}{rand}
    \SetCommentSty{mycommfont}
    \SetKwRepeat{Repeat}{do}{while}
    \SetKw{In}{\textbf{in}}
    \SetKw{And}{\textbf{and}}
    \SetKw{Is}{\textbf{is}}
    
    \Output{ {\em query}}	
    \Fn{\GenQuery{}}{
        {\em query} $\leftarrow$ \EmptyQuery{}\;\label{alg_init_start}
        {\em qc} $\leftarrow$ \{\}; \tcp{query context}
        {\em gs} $\leftarrow$ \{\}; \tcp{graph schema}\label{alg_init_end}
        \Repeat{\rand{} < \P \And \length{query} < \L} 
        {\label{alg_append_start}
            {\em clause}, {\em qc}, {\em gs} $\leftarrow$ \GenClause{qc, gs, \any}\;\label{alg_update_state_query}
            \append{query, clause}\;
        }\label{alg_append_end}
        \textbf{return} {\em query}\;\label{alg_return_query}
    }
    \Fn{\GenClause{qc, gs, type}}{
        \If{type = \any}{ \label{alg_gen_type_start}
           {\em type} = \RandClauseType{}\;
        }\label{alg_gen_type_end}
        \tcp{initialize the clause accordingly}
        {\em clause}, {\em qc}, {\em gs} $\leftarrow$ \InitializeClause{qc, gs, type}\;\label{alg_init_clause}
        \ForEach{subclause, subclause\_type \In clause}{ \label{alg_subclause_start}
            {\em subclause}, {\em qc}, {\em gs} $\leftarrow$ \GenClause{qc, gs, subclause\_type}\;
        }\label{alg_subclause_end}
        
        \tcp{clean out-of-scope query context}
        \CleanQC{qc}\; \label{alg_clean_qc}
        \textbf{return} {\em clause}, {\em qc}, {\em gs}\;\label{alg_return_clause}
    }
    
  \caption{State-Aware Query Generation}
  \label{alg_query_generation}
\end{algorithm}

\vspace{\midspace}
\noindent
\textbf{Algorithm.}
Algorithm~\ref{alg_query_generation} shows the procedure of our state-aware query generation. Our approach does not need any input.
Initially, it constructs an empty query that contains no clause, and initializes query context \emph{qc} and graph schema \emph{gs} to empty sets (line~\ref{alg_init_start}-\ref{alg_init_end}). Then, the approach incrementally appends the query with a newly generated clause (line~\ref{alg_append_start}-\ref{alg_append_end}). Each time a new clause is generated, the query context and graph schema may be updated (line~\ref{alg_update_state_query}). 
Our approach decides whether to append the query with more clauses based on a certain probability P. If the length of the query reaches the limit L$_{max}$, the approach stops appending to the query (line~\ref{alg_append_end}). In the end, the generated query is returned (line~\ref{alg_return_query}).


To generate a clause, our approach references the query context and graph schema. If the clause type is not specified, the approach randomly chooses a clause type (line~\ref{alg_gen_type_start}-\ref{alg_gen_type_end}). According to the elements in query context, graph schema, and the specified clause type, our approach randomly initializes a clause (line~\ref{alg_init_clause}). Initialization for different clauses can vary (\emph{e.g.}, the procedures for \cypher|CALL| and \cypher|CREATE| are different), but it generally includes the procedure of determining the number of components needed by the clause and their types, generating corresponding expressions for these components, and analyzing the impacts of the generated clause. If the generated clause contains subclauses, the approach will recursively call \cypher{GenClause()} to generate subclauses with specified types (line~\ref{alg_subclause_start}-\ref{alg_subclause_end}).
After the clause is generated, the approach cleans up the query context if necessary (line~\ref{alg_clean_qc}), such as removing out-of-scope variables at the end of a \cypher{CALL} clause. In the end, the generated clause, the updated query context, and the graph schema are returned (line~\ref{alg_return_clause}).

\begin{figure}
\centering\includegraphics[width=0.95\linewidth]{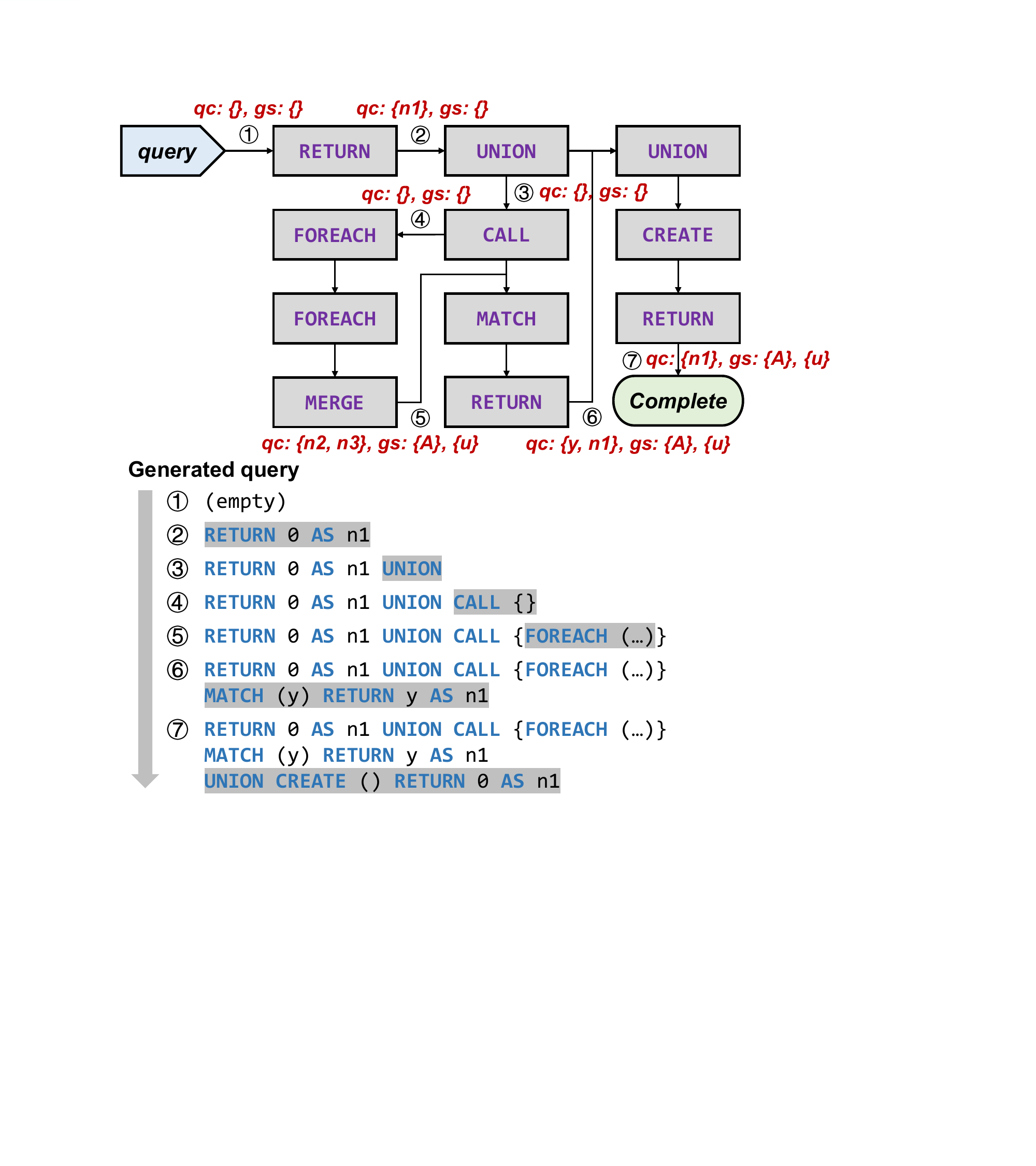}
	\caption{Generation for the original query in  Figure~\ref{fig_neo4j_bug_intro}.} 
	\label{fig_query_generation}
    \vspace{-5pt}
\end{figure}

\subsubsection{Illustrative Example}
We show the effectiveness of our generation using a complex query example, the original query in Figure~\ref{fig_neo4j_bug_intro}. As shown in Figure~\ref{fig_query_generation}, initially, the query is empty, and both query context and graph schema contain nothing (step \ding{192}). The approach randomly determines that the first clause is a \cypher{RETURN} clause with an integer \cypher{0} as the returned value \cypher{n1}. Then, the approach updates the query context by adding \cypher{n1} with its type and scope (step \ding{193}).
The second clause is a \cypher{UNION} clause. After constructing the \cypher{UNION}, the approach cleans the query context as the clause after the \cypher{UNION} cannot reference the variables defined before the \cypher{UNION} (step \ding{194}). Subsequently, the approach randomly decides to generate a \cypher{CALL} clause with a subclause (step \ding{195}). The subclause is recursively generated as a \cypher{FOREACH} clause, defining \cypher{n2} in its loop condition. The \cypher{FOREACH} also has a subclause, which is another \cypher{FOREACH} clause, defining \cypher{n3}. The last \cypher{FOREACH} has a \cypher{MERGE} as its subclause. Our approach records the query context (\emph{i.e.}, \cypher{n2}, \cypher{n3} defined by the two \cypher{FOREACH} clauses)
and graph schema (\emph{i.e.}, label \cypher{A} and property \cypher{u} referenced by the \cypher{MERGE}) updated in this recursive generation process (step \ding{196}). After completing generation of the \cypher{FOREACH} clauses, the approach removes \cypher{n2} and \cypher{n3} from query context as they are scoped to only the \cypher{FOREACH} clauses. Finishing the \cypher{CALL}, our approach continues to randomly generate clauses, including a \cypher{MATCH} defining \cypher{y}, and a \cypher{RETURN} referencing \cypher{y} as \cypher{n1}, and update the query context by adding \cypher{y} and \cypher{n1} (step \ding{197}). Finally, our approach randomly generates a \cypher{UNION} followed by a \cypher{CREATE} and a \cypher{RETURN}. The \cypher{UNION} cleans all the contents of the query context, after which \cypher{RETURN} adds a new variable \cypher{n1} (step \ding{198}).

\subsection{Fine-Grained Query Transformation}

To enable general query validation, we construct test oracles based on the fine-grained units of queries: clauses and expressions.
Each query consists of a chain of clauses, which can be transformed by modifying the clauses or prepending them with new clauses.
Each clause may include expressions, on which more transformations can be applied by using algebraic identities or Cypher features.

\begin{table}[t]
        \caption{Clause-level transformation rules. (1) \textbf{T.S.}: transformation strategies including Dead Code (DC), Unused Context (UC), Redundant Write (RW), and Inconsequential Supersession (IS); (2) \cypher{<literal>}: any atomic expression; (3) \cypher{<mirror(x)>}: a mirrored version of \cypher+x+ (\emph{e.g.}, \cypher+a->b+ $\Rightarrow$ \cypher+b<-a+); (4) \cypher{<clauses>}: a random number of random clauses; (5) \cypher{<new pattern>}: a graph pattern that involves at least one previously unused label or property}
	\label{tbl_clause_transformations}
        \setlength\tabcolsep{1.9pt}
	\noindent{\small
		\begin{center}
            \begin{tabular}{cccl}
            \toprule
            \textbf{Type} & \textbf{Original} & \textbf{T.S.} & \textbf{Transformed} \\
            \midrule
            \multirow{7}{*}{\shortstack{Read-\\ Only}} & \multirow{4}{*}{\centering\cypher+C+} & DC & \cypher+CALL {MATCH x WHERE false <clauses>} C+ \\
            \cline{3-4}
            & & DC & \cypher+CALL {MATCH <new pattern> <clauses>} C+ \\
            \cline{3-4}
            & & IS & \cypher+OPTIONAL MATCH <new pattern> C+ \\
            \cline{3-4}
            & & UC & \cypher+CALL {WITH * RETURN x AS i} C+ \\
            \cline{3-4}
            & & UC & \cypher+UNWIND <literal> AS i C+ \\
            \cline{2-4}
            & \cypher+WITH x+ & UC & \cypher+WITH x, y AS i+ \\
            \cline{2-4}
            & \multirow{2}{*}{\centering\cypher+MATCH x+} & IS & \cypher+MATCH x WHERE true+ \\
            \cline{3-4}
            & & IS & \cypher+MATCH <mirror(x)>+ \\
            \midrule
            \multirow{7}{*}{\shortstack{Write/\\Project}} & \multirow{4}{*}{\centering\cypher+C+} & DC & \cypher+FOREACH (i IN [] |<clauses>) C+ \\
            \cline{3-4}
            & & RW & \cypher+DELETE <deleted var> C+ \\
            \cline{3-4}
            & & RW & \cypher+CREATE x DELETE x C+ \\
            \cline{3-4}
            & & RW & \cypher@SET x += {} C@ \\
            \cline{2-4}
            & \cypher+REMOVE x+ & RW & \cypher+REMOVE x, x+ \\
            \cline{2-4}
            & \multirow{2}{*}{\centering\cypher+DELETE x+} & RW & \cypher+DELETE x, x+ \\
            \cline{3-4}
            & & IS & \cypher+DETACH DELETE x+ \\
            
            \midrule
            \centering Any & \multirow{1}{*}{\centering\cypher{C}} & IS & \cypher+WITH * C+ \\
            \bottomrule
            \end{tabular}
            
        \end{center}
    }

\end{table}


\subsubsection{Clause-Level Transformations}

These transformations manipulate queries by either adding new clauses, or switching out clauses with equivalent counterparts. 
Table~\ref{tbl_clause_transformations} lists the transformation rules. We design different rules for reading clauses and writing/projecting clauses because the Cypher specification limits the kinds of clauses that can be used before specific clauses.
The transformations consider state information to avoid semantic invalidity (\emph{e.g.}, repeatedly defining the same variables).

We introduce four strategies to achieve semantic-preserving transformation for clauses: \emph{Dead Code (DC)}, \emph{Unused Context (UC)}, \emph{Redundant Write (RW)}, and \emph{Inconsequential Supersession (IS)}.
\emph{Dead Code} expands the target clause by introducing additional clauses that, due to Cypher's way of handling queries, will not be executed. 
We implement three rules for this strategy. 
Two rules transform read clauses by adding a \cypher+CALL+ clause with a \cypher{MATCH}. The \cypher{MATCH} clause is designed to return no rows, leading to no rows for the clauses following the \cypher{MATCH} to ingest. As a result, the subsequent clauses in the \cypher{CALL} clauses will not be executed. One rule transforms write clauses by inserting a \cypher{FOREACH} before them. The \cypher{FOREACH} iterates the elements in a given array and invokes clauses to operate on the iterated elements. Because the array is empty, no iterations are performed by the \cypher{FOREACH}. Therefore, the invoked clauses in the \cypher{FOREACH} will not be executed. For example, in Figure~\ref{fig_neo4j_bug_intro}, our approach transforms the first \cypher{RETURN} clause of the original query by adding a dead \cypher{FOREACH}, which preserves the semantics of the \cypher{RETURN} and thus should not affect the final results. However, the results of the transformed query change, indicating that a bug was triggered.

\emph{Unused Context} expands a clause by involving unused query context. Specifically, it defines new variables that are not referenced by subsequent parts of the query. \emph{Redundant Write} is straightforward. It expands a clause by deleting a previously deleted node, removing the same property multiple times, or creating and immediately deleting a subgraph.
Lastly, \emph{Inconsequential Supersession} expands a clause by adding inconsequential operations (\emph{e.g.}, adding \cypher{WHERE true} to a \cypher{MATCH} clause) or modifying the formats of the clauses (\emph{e.g.}, reversing the graph pattern of a \cypher{MATCH} clause).


Clause-level transformations have the ability to influence the query context through manipulating variables, as well as changing the database state by modifying graph elements or affecting control flow. By doing so, \sys can effectively force the GDBMSs to execute system logic that was not covered by the original queries, and thus validate the query results by checking whether the different executed logic produces the same results.

\begin{table}[t]
\caption{Expression-level transformation rules. (1) \textbf{T.S.}: transformation strategies including Mathematical Identity (MI) and Cypher Feature (CF); (2) \cypher{P'}: randomly generated boolean expression; (3) \cypher{E'}: randomly generated expression}
	\label{tbl_expression_transformations}
        \setlength\tabcolsep{3.3pt}
	\noindent{\small
		\begin{center}
			\begin{tabular}{cccl}
                \toprule
                \textbf{Type} & \textbf{Initial} & \textbf{T.S.} & \textbf{Transformed} \\
                \midrule
                \multirow{11}{*}{\centering Bool} & \multirow{3}{*}{\centering\cypher+P+} & MI & \cypher+NOT (NOT P)+ \\
                & & MI & \cypher+P OR false+ \\
                & & MI & \cypher+P AND true+ \\
                \cline{2-4}
                & \cypher+P OR Q+ & MI & \cypher+NOT ((NOT P) AND (NOT Q))+ \\
                \cline{2-4}
                & \cypher+P AND Q+ & MI & \cypher+NOT ((NOT P) OR (NOT Q))+  \\
                \cline{2-4}
                & \multirow{2}{*}{\centering\cypher+true+} & MI & \cypher+true OR P'+ \\
                & & MI & \cypher+P' OR (NOT P') OR (P' IS NULL)+ \\
                \cline{2-4}
                & \multirow{2}{*}{\centering\cypher+false+} & MI & \cypher+false AND P'+\\
                & & MI & \cypher|P' AND (NOT P') AND (P' IS NOT NULL)| \\
                \cline{2-4}
                & \cypher+NULL+ & MI & \cypher+NULL {XOR|=|<>|<|>|>=|<=} P'+ \\
                \midrule
                \multirow{5}{*}{\centering Num} & \multirow{3}{*}{\centering\cypher+x+} & MI & \cypher+-(-x)+ \\
                \cline{3-4}
                & & MI & \cypher@x {+|-} 0@ \\
                \cline{3-4}
                & & MI & \cypher+x {*|/|^} 1+ \\
                \cline{2-4}
                & \cypher+0+ & MI & \cypher+0 * x+ \\
                \cline{2-4}
                & \cypher+NULL+ & MI & \cypher@NULL {+|-|*|/} x@ \\
                \midrule
                Str & \cypher+S+ & MI & \cypher@S + ""@ \\
                \midrule
                List & \cypher+L+ & MI & \cypher@L + []@ \\
                \midrule
                \multirow{3}{*}{Any} & \cypher+E IS NULL+ & MI & \cypher+NOT (E IS NOT NULL)+ \\
                \cline{2-4}
                & \multirow{2}{*}{\cypher+E+} & CF & \cypher+CASE WHEN true THEN E ELSE E' END+ \\

                & & CF & \cypher+CASE WHEN false THEN E' ELSE E END+ \\
                \bottomrule
            \end{tabular}
        \end{center}
    }
\end{table}

\subsubsection{Expression-Level Transformations}

These transformations operate on expressions within Cypher clauses. Although they do not modify the state of the database or query, they still force the GDBMS to evaluate more complex expressions.
Table~\ref{tbl_expression_transformations} lists our expression-level transformations.

Our expression-level transformations involve two strategies: Mathematical Identity (MI) and Cypher Feature (CF). Mathematical Identity leverages mathematical laws (\emph{e.g.}, De Morgan's laws for boolean con- and disjunctions, and double negations) to transform the expressions while preserving their semantics. Cypher Feature leverages the features of the Cypher language. For example, \sys can transform an arbitrary expression into a \cypher{CASE} expression, where the branch condition is determined to be \cypher{true}, and the corresponding value of the true branch is the original expression. When transforming expressions, \sys sometimes needs to involve new components in the transformed expressions (\emph{e.g.}, the \cypher{E'} in the transformed \cypher{CASE} expressions). In this case, \sys constructs these components by referencing the available variables recorded in the corresponding query context.

\section{Implementation}
\label{sec_implementation}

We implement our approach as a fully automatic GDBMS testing framework, \sys.
The overall codebase of \sys consists of 10k lines of Go code. The source code of \sys is available for review on an anonymized repository\footnote{\url{https://github.com/Anon10214/dinkel}} and will be public concurrent with the publication of the paper.
The following discusses important implementation details.

\vspace{\midspace}
\noindent
\textbf{Supported Clauses in Query Generation.}
\sys supports all non-administrative clauses in the tested GDBMSs. For example, in Neo4j, 21 out of the total 25 clauses are non-administrative and thereby supported.
Administrative clauses, such as \cypher{SHOW FUNCTIONS} and \cypher{USE}, are not supported because they are not related to query processing logic.
For example, 
\cypher{SHOW FUNCTIONS} is designed to list all available functions and cannot be combined with other clauses. 
\cypher{USE} is used to switch the manipulated databases, which is not necessary as all databases are empty on initialization and they function in the same way.

As shown in Table~\ref{tbl_supported_clause}, \sys supports more Cypher clauses than existing approaches~\cite{jiang2023icse, matteo2023issta, ziyue2023issta, zhuang2023vldb, qiuyang2024grev}. 
None of the existing approaches support \cypher|FOREACH| clauses, because this clause complicates control flow and can significantly modify graph states during query execution. \cypher|UNION|, \cypher|EXISTS|, and \cypher|COUNT| are not supported either, as these clauses invoke subqueries that can inherit and change the graph states of the main query. 
Lacking a systematic model for handling graph state changes, existing approaches are limited in generating complex queries using these clauses.
Moreover, existing approaches support validation for only queries following specific patterns, while many clauses complicate the semantics of Cypher queries, making them unable to be validated by existing approaches.




\begin{table}[t]
	\caption{Cypher clauses supported by existing approaches}
	\label{tbl_supported_clause}
	\noindent{
 \small
		\begin{center}
  \setlength{\tabcolsep}{3.1pt}
			\begin{tabular}{lccccc}
			\toprule
Clause & GDsmith & GDBMeter & GraphGenie & GAMERA & Dinkel \\
\midrule
\cypher|MATCH|   & \filledcircle & \filledcircle & \filledcircle & \filledcircle & \filledcircle  \\
\cypher|CREATE|  & \filledcircle & \filledcircle & \filledcircle & \filledcircle & \filledcircle \\
\cypher|MERGE|   & \emptycircle & \emptycircle & \emptycircle & \emptycircle & \filledcircle \\
\cypher|DELETE|  & \emptycircle & \filledcircle & \emptycircle & \emptycircle & \filledcircle \\
\cypher|REMOVE|  & \emptycircle & \filledcircle & \emptycircle & \emptycircle & \filledcircle \\
\cypher|SET|     & \emptycircle & \filledcircle & \emptycircle & \emptycircle & \filledcircle \\
\cypher|UNWIND|  & \filledcircle & \emptycircle & \filledcircle & \filledcircle & \filledcircle \\
\cypher|WITH|    & \filledcircle & \emptycircle & \filledcircle & \filledcircle & \filledcircle \\
\cypher|RETURN|  & \filledcircle & \filledcircle & \filledcircle & \filledcircle & \filledcircle \\
\cypher|CALL|    & \emptycircle & \emptycircle & \filledcircle & \emptycircle & \filledcircle \\
\cypher|FOREACH| & \emptycircle & \emptycircle & \emptycircle & \emptycircle & \filledcircle \\
\cypher|UNION|   & \emptycircle & \emptycircle & \emptycircle & \emptycircle & \filledcircle \\
\cypher|EXISTS|  & \emptycircle & \emptycircle & \emptycircle & \emptycircle & \filledcircle \\
\cypher|COUNT|   & \emptycircle & \emptycircle & \emptycircle & \emptycircle & \filledcircle \\
\bottomrule
\end{tabular}
\end{center}}
\end{table}

\vspace{\midspace}
\noindent
\textbf{Bug Detection.}
\sys detects logic bugs via its fine-grained query transformation. Besides logic bugs, \sys applies some methods to identify internal errors and memory bugs in GDBMSs. 
Specifically, 
for each executed query, \sys checks the messages returned from the GDBMS. If the message indicates an internal error in the GDBMS (\emph{e.g.}, the error in Figure~\ref{fig:query_largest_size}), a bug is identified.
We also enable all embedded assertions in the tested GDBMSs to catch assertion failures.
In addition, we enable ASan~\cite{ASan} for the GDBMSs developed in C/C++ to identify memory bugs triggered by our generated queries.

\vspace{\midspace}
\noindent
\textbf{Query Reduction.}
No existing tool is available for reducing Cypher queries, which makes it difficult for developers to minimize bug-triggering queries and investigate bugs. To address this problem, 
we implemented \sys with an automatic query reduction method. The core idea of this method is to reduce queries clause by clause. For each clause, \sys tries to delete it. If the query without the clause still triggers the bug, the clause will be permanently removed. Otherwise, the deleted clause will be recovered, and \sys goes on to try to replace the clause with an alternative clause if possible. For example, \sys can try to replace the \cypher{CALL} clause in Figure~\ref{fig_neo4j_bug_intro} with its subclause \cypher{CREATE}. 
If some clauses are successfully deleted or replaced in one try, \sys will restart the reduction process for the reduced queries. The process stops when no clause in the query can be further reduced.

\section{Evaluation}

Our evaluation aims to answer the following questions:

\begin{enumerate}[start=1,label={\bfseries Q\arabic*},font=\itshape, topsep=0pt, leftmargin=*]

\item Can \sys find real bugs in widely used and extensively tested GDBMSs? (Section~\ref{sec_bug_detection})

\item How complex and valid are the queries generated by \sys? (Section~\ref{sec_query_generation})

\item How do different techniques contribute
to the effectiveness of \sys? (Section~\ref{sec_sensitivity_analysis})

\item Can \sys outperform state-of-the-art GDBMS testing approaches? (Section~\ref{sec_comparison})

\end{enumerate}

\subsection{Experimental Setup}

We evaluated \sys on Neo4j~\cite{neo4j-repo}, RedisGraph~\cite{redisgraph-repo}, and Memgraph~\cite{memgraph-repo}, which are popular and extensively tested by existing approaches~\cite{ziyue2023issta, jiang2023icse, matteo2023issta, zhuang2023vldb, qiuyang2024grev}.
We evaluate \sys on the latest GDBMS versions. During testing, if the tested GDBMSs are updated (\emph{e.g.}, a new version is released), we set up new \sys instances to test the updated versions. Specifically, we test Neo4j from version 5.6.0, RedisGraph from 2.12.0, and Memgraph from 2.7.0. RedisGraph is no longer maintained after 2.12.10, and thus we move to test its successor, FalkorDB~\cite{falkordb-repo}, from its first release version 4.0.0.
We count the bugs of RedisGraph and FalkorDB together as they share the majority of their code base.
To show its effectiveness for black-box testing GDBMSs, we evaluate \sys on the Enterprise version of Neo4j from 5.6.0 onward. Each time we implement new features on \sys, we stop and restart the testing. Overall, the testing campaign intermittently lasted 9 months. The evaluation was performed on a machine running Ubuntu 20.04,
with a 64-core AMD EPYC 7742 processor running at 2.25GHz and 256GB of RAM.

\subsection{Bug Detection}
\label{sec_bug_detection}

Table~\ref{tbl_bug_status} shows the status of the bugs found by \sys. In total, \sys found \bugsfound unique bugs, including 51 bugs in Neo4j, 62 in RedisGraph, and 14 in Memgraph. Among these bugs, \bugsconfirmed are confirmed, and \bugsfixed are fixed. None of the \bugsfound bugs are duplicates. As mature GDBMSs, Neo4j, RedisGraph, and Memgraph have been extensively tested in both industry and academia. The significant number of new bugs demonstrates the powerful bug-finding ability of \sys. We will publish all our \bugsfound bug reports afterward with our artifact.

\vspace{\midspace}
\noindent
\textbf{Bug Classification}. We classify the bugs found by \sys into three categories according to their manifestation:
\begin{itemize}[leftmargin=*]

\item \textit{Logic bugs.} 
The tested GDBMSs produce incorrect results for specific queries. These bugs were exposed because they lead to discrepancies between the results of the original queries and the transformed queries generated by \sys.

\item \textit{Internal errors.} 
The tested GDBMSs unexpectedly throw exceptions or errors when processing syntactically and semantically valid queries. The error messages can indicate the inconsistency of the internal execution status.

\item \textit{Crashes.} 
The queries cause GDBMSs to crash due to assertion failures or memory corruption.

\end{itemize}

\begin{table}[t]
	\caption{Status of the bugs found by \sys{}}
	\label{tbl_bug_status}
	\noindent{\small
		\begin{center}
			\begin{tabular}{@{}lrrr@{}}
			\toprule
\textbf{GDBMS} & \textbf{Reported} & \textbf{Confirmed} & \textbf{Fixed} \\
\midrule
Neo4j & 51 & 51 & 49  \\
RedisGraph & 62 & 48 & 29  \\
Memgraph & 14 & 14 & 6   \\
\cmidrule{1-4}
\textbf{Total} & \bugsfound & \bugsconfirmed & \bugsfixed  \\
\bottomrule
\end{tabular}
\end{center}}
\end{table}

\begin{table}[t]
	\caption{Classifying the found bugs}
	\label{tbl_bug_classification}
	\noindent{\small
		\begin{center}
			\begin{tabular}{@{}lrrr@{}}
			\toprule
\textbf{GDBMS} & \textbf{Logic Bug} & \textbf{Internal Error} & \textbf{Crash}\\
\midrule
Neo4j & 13 & 38 & 0\\
RedisGraph & 18 & 10 & 34\\
Memgraph & 2 & 8 & 4\\
\cmidrule{1-4}
\textbf{Total} & 33 & 56 & 38\\
\bottomrule
\end{tabular}
\end{center}}
\end{table}

Table~\ref{tbl_bug_classification} shows the classification results. Among the \bugsfound bugs \sys found, 33 are logic bugs, 56 cause internal errors, and 38 result in GDBMS crashes.
Note that Neo4j is implemented in Java and has exception handling for unexpected memory errors. Thus, Neo4j does not crash, but rather responds with exception information when triggering memory corruption. Therefore, \sys found no crashes in Neo4j, but 38 internal errors. 
Among the 38 crashes found in RedisGraph and Memgraph, 29 are caused by memory corruptions, and 9 are caused by assertion failures. These results demonstrate that \sys can comprehensively test GDBMSs by finding various bugs.

Figure~\ref{fig:query_largest_size} shows a bug that triggers a Neo4j internal error.
This query constructs an empty array \cypher|[]| and invokes a subquery using a \cypher|CALL|. The subquery also constructs an empty array. It then invokes two \cypher|UNWIND| clauses, which iterate over each element in the operated array. For each iterated element, the \cypher|UNWIND| executes the subsequent clause under the context of this element. For example, the array \cypher|[x]| used by the second \cypher|UNWIND| references the variable \cypher|x|, which is \cypher|0| when the first \cypher|UNWIND| iterates over the item \cypher|0| in the array \cypher|[0]|. 
After the \cypher|CALL|, the query utilizes \cypher|FOREACH|, whose execution depends on the query context of the \cypher|CALL|. To optimize query execution, Neo4j tries to flatten the \cypher|FOREACH| loop. However, such optimization does not work well when the loop involves update operations (\emph{i.e.}, \cypher|MERGE|) under complicated contexts (\emph{i.e.}, the contexts produced by \cypher|CALL|). The improper optimization corrupts the internal data structures of Neo4j, \textit{Eagers}, which are the production of another optimization, \textit{Eagerness analysis}. In the end, an internal error is triggered when Neo4j tries to access the corrupted Eagers. To fix this bug, Neo4j developers modify both the Eagerness analysis and the flattening strategy for \cypher|FOREACH| clauses to ensure they work consistently.

\vspace{\midspace}
\noindent
\textbf{Affected GDBMS Components.} We investigated 59 of the \bugsfixed fixed bugs, where we were able to analyze the fix patches to identify the GDBMS components affected by the bugs accurately. 
The 19 fix patches in the Enterprise version of Neo4j are confidential and thus not included.
The 6 fixed bugs in Memgraph are also not included because Memgraph developers mix the fixes into their regular development commits, where we cannot accurately analyze the affected components.
Table~\ref{tbl_bug_components} shows the results.
Among the 59 fixed bugs, 29 affect the parsers of GDBMSs, 25 affect the planners, and 20 affect the executor. 
Interestingly, 29 bugs (5 logic bugs) affect two components, and 12 bugs (1 logic bug) affect all the components. For example, the bug shown in Figure~\ref{fig_neo4j_bug_RW_dependencies} affects both the planner and executor of Neo4j. In total, 40 bugs (8 logic bugs) affect either the planner, the executor, or both.  The 6 logic bugs affecting the executor also affect the planner.
These results demonstrate that (1) \sys can find bugs in various GDBMS components; and (2) \sys can find bugs in the deep logic of GDBMSs, considering 67.8\% (40/59) of fixed bugs are related to the planners or executors. 



\begin{figure}[t]
    \begin{lstlisting}[language=cypher]
WITH [] AS n0 ORDER BY null
CALL {
  WITH [] AS n1 ORDER BY null
  UNWIND [0] AS x
  UNWIND [x] AS n2
  RETURN n2 AS n3
} FOREACH (n4 IN null | MERGE ({key:n3}))
    \end{lstlisting}
    \caption{A query triggering a Neo4j internal error---Neo4j-\\Error: ExecutionFailed (arraycopy: last destination index 6 out of bounds for object array[5]).}
    \label{fig:query_largest_size}
\end{figure}

\begin{table}[t]
	\caption{GDBMS components affected by bugs. The number in the parentheses is for the logic bugs}
	\label{tbl_bug_components}
	\noindent{\small
		\begin{center}
			\begin{tabular}{@{}lrrr@{}}
			\toprule
\textbf{GDBMS} & \textbf{Parser} & \textbf{Planner} & \textbf{Executor}\\
\midrule
Neo4j & 18 (0) & 18 (5) & 10 (3)\\
RedisGraph & 15 (1) & 15 (3) & 19 (3)\\
\cmidrule{1-4}
\textbf{Total} & 23 (1) & 23 (8) & 29 (6)\\
\bottomrule
\end{tabular}
\end{center}}
\end{table}

\begin{figure}
    \begin{center}
        \includegraphics[width=0.92\linewidth]{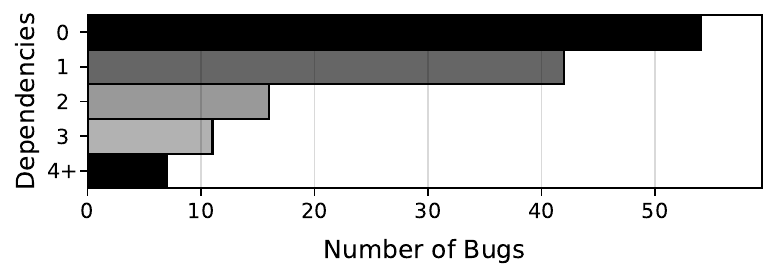}
    \end{center}
    \caption{\label{fig:bug-dependencies} Data dependencies of bug-triggering queries. }
\end{figure}

\vspace{\midspace}
\noindent
\textbf{Data Dependencies for Triggering Bugs.} 
To convey the data dependencies required to trigger bugs, we analyze dependencies in the \bugsfound bug-triggering queries.
The results are shown in Figure~\ref{fig:bug-dependencies}.
Over half of the bugs (58\%) require at least one data dependency. Specifically, 42 bug-triggering queries contain one data dependency, 16 contain two, 11 contain three, and 7 contain four or more. For example, in Figure~\ref{fig:assertion_failure_redisgraph}, the query contains 8 data dependencies, as illustrated in Figure~\ref{fig:datadependenciesexample}. Among the 8 dependencies, 6 are involved in query context (\emph{i.e.}, the variables \lstinline|x| and \lstinline|y| assigned to a node and a relationship), and 2 are involved in graph schema (\emph{i.e.}, the label \lstinline|A|). These results demonstrate that some GDBMS bugs can be triggered only when the queries contain multiple data dependencies, and \sys can effectively find these bugs.

\begin{figure}
    \begin{center}
        \colorlet{bblue}{blue!70!black}
\colorlet{bbblue}{bblue!20}
\colorlet{ppurple}{purple!70!black}
\colorlet{pppurple}{ppurple!20}
\colorlet{ggreen}{green!50!black}
\colorlet{gggreen}{ggreen!20}

\DeclareRobustCommand{\x}{\sethlcolor{bbblue}\hl{x}}
\DeclareRobustCommand{\A}{\sethlcolor{pppurple}\hl{A}}
\DeclareRobustCommand{\y}{\sethlcolor{gggreen}\hl{y}}

\ttfamily

\scalebox{0.93}{
\begin{tikzpicture}[
        arrow/.style={-{Latex[length=1mm,width=2mm]}},
        link0/.style={arrow, to path={-- ++(0,2mm) -| (\tikztotarget)}},
        link1/.style={arrow, to path={-- ++(0,4mm) -| (\tikztotarget)}},
        dashedd/.style={dash pattern={on 3pt off 2pt}},
        dasheddott/.style={dash pattern={on 5pt off 1pt on 1pt off 1pt}}
    ]
    \node (a0) {MERGE (};
    \node[right=-7pt of a0] (a1) {\x};
    \node[right=-7pt of a1] (a2) {)<-[:};
    \node[right=-7pt of a2] (a3) {\A};
    \node[right=-7pt of a3] (a4) {]-(};
    \node[right=-7pt of a4] (a5) {\x};
    \node[right=-7pt of a5] (a6) {)<-[:};
    \node[right=-7pt of a6] (a7) {\A};
    \node[right=-7pt of a7] (a8) {]-(};
    \node[right=-7pt of a8] (a9) {\x};
    \node[right=-7pt of a9] (a10) {)<-[};
    \node[right=-7pt of a10] (a11) {\y};
    \node[right=-7pt of a11] (a12) {:};
    \node[right=-7pt of a12] (a13) {\A};
    \node[right=-7pt of a13] (a14) {]->(};
    \node[right=-7pt of a14] (a15) {\x};
    \node[right=-7pt of a15] (a16) {)};

    \node[below=11pt of a0.west, anchor=west] (b0) {DELETE};
    \node[right=-3pt of b0] (b1) {\y};
    
    \node[below=11pt of b0.west, anchor=west] (c0) {CREATE (};
    \node[right=-7pt of c0] (c1) {\x};
    \node[right=-7pt of c1] (c2) {)<-[:};
    \node[right=-7pt of c2] (c3) {B};
    \node[right=-7pt of c3] (c4) {]-()};
    
    \node[below=11pt of c0.west, anchor=west] (d0) {DELETE};
    \node[right=-3pt of d0] (d1) {\y};

    \draw[-, thick, bblue] (a1.north)++(0,-1mm) edge[link0] ([yshift=-1mm]a5.north);
    \draw[-, thick, bblue] (a1.north)++(0,-1mm) edge[link0] ([yshift=-1mm]a9.north);
    \draw[-, thick, bblue] (a1.north)++(0,-1mm) edge[link0] ([yshift=-1mm]a15.north);
    \draw[-, thick, bblue] (a1.north)++(0, 1mm) edge[arrow, to path={-- ++(-1.9mm,0) |- (\tikztotarget)}] ([xshift=1mm]c1.west);
    
    \draw[-, thick, ppurple] (a3.north)++(0,-1mm) edge[dashedd, link1] ([yshift=-1mm]a7.north);
    \draw[-, thick, ppurple] (a3.north)++(0,-1mm) edge[dashedd, link1] ([yshift=-1mm]a13.north);
    
    \draw[arrow, thick, ggreen, dasheddott] (a11.south)++(0,1mm) |- ([xshift=-1mm]b1.east);
    \draw[arrow, thick, ggreen, dasheddott] (a11.south)++(0,1mm) |- ([xshift=-1mm]d1.east);
\end{tikzpicture}
}
    \end{center}
    \caption{\label{fig:datadependenciesexample} Data dependencies within the query in Figure~\ref{fig:assertion_failure_redisgraph}. }
\end{figure}

\begin{figure}
    \begin{center}
        \includegraphics[width=0.85\linewidth]{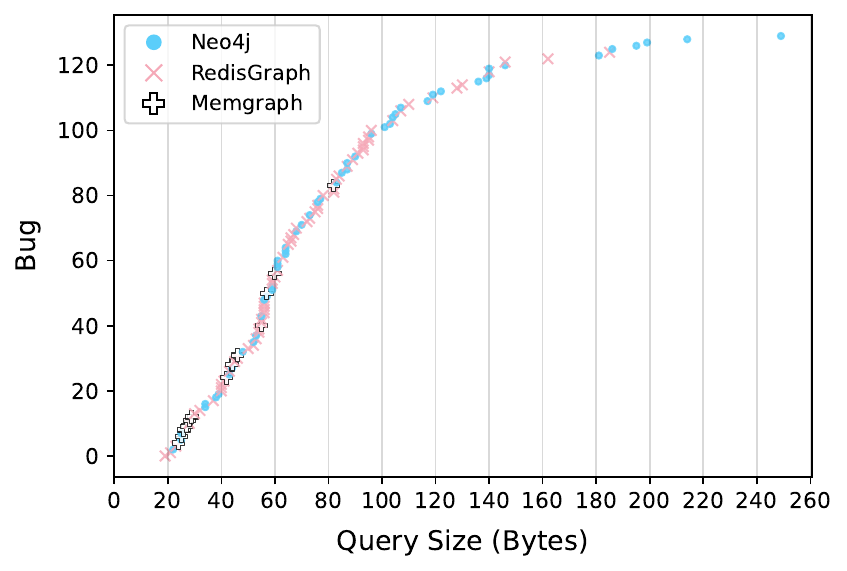}
    \end{center}
    \caption{\label{fig:bug-sizes} Query sizes of bug-triggering queries. }
\end{figure}

\vspace{\midspace}
\noindent
\textbf{Size of Bug-Triggering Queries.} 
Figure~\ref{fig:bug-sizes} shows the size of the \bugsfound bug-triggering queries. All these queries have been reduced using the methods mentioned in Section~\ref{sec_implementation}. 112 bugs can be triggered by queries whose size is less than 120 bytes. 
The bug-triggering query shown in Figure~\ref{fig:assertion_failure_redisgraph} is an example of such a query.
As query size increases from 0 to 120 bytes, the number of triggered bugs increases almost linearly. The bug-triggering query with the largest size is 249 bytes, which triggers an internal error in Neo4j and was fixed by their developers.



\vspace{\midspace}
\noindent
\textbf{Bug Importance.}
Neo4j, the GDBMS we focus on, provides Community and Enterprise versions. The bugs in the Enterprise version are critical because this version is commonly deployed on commercial applications. Among the 51 Neo4j bugs \sys found, 32 bugs can be triggered in both Enterprise and Community versions, and 19 bugs can be triggered in only the Enterprise version. RedisGraph and Memgraph did not provide severity information about the reported bugs, but we noticed that most bugs were fixed, indicating that the bugs are non-trivial. 
Some developers express their appreciation for our effort in finding bugs in their GDBMSs. Particularly, Neo4j provides very positive feedback:

\begin{tcolorbox}[colback=mygray, 
                  rounded corners,
left=1.5mm, right=1.5mm, top=1.5mm, bottom=1.5mm,
boxrule=0.2mm]
My colleague has told me that you've been busy creating GitHub issues and have contributed to improving Neo4j! Thanks a lot for that! In the name of Neo4j, I would like to send you some swag [...]


\end{tcolorbox}

\subsection{Query Generation}
\label{sec_query_generation}

To understand the complexity and validity of the queries constructed by \sys, we ran \sys on each tested GDBMS for 48 hours and collected all generated queries 
(transformed queries are not included). 
Table~\ref{tbl_query_gen} shows the statistical results.

\vspace{\midspace}
\noindent
\textbf{Query Complexity}. 
On average, \sys generates 655k queries for each GDBMS, with 22.62M clauses constructed.
The average number of clauses of each query is 34.54, and the number of data dependencies is 26.36.
Because of the large number of clauses and dependencies, the queries generated by \sys are typically large. The average query size is 1153.3 bytes. 
These results demonstrate that \sys can effectively generate complex queries, which are large and contain multiple clauses with complicated data dependencies. 
The generated queries are more complex than the bug-triggering queries discussed in Section~\ref{sec_bug_detection}, because all the bug-triggering queries are reduced, while the generated queries inevitably contain many redundant parts~\cite{zuming2023osdi, David2020ecoop, C-Reduce}.


\begin{table}[t]
	\caption{Queries generated by \sys{} within 48 hours. \textbf{Gen.}: the number of generated queries; \textbf{Valid}: the number of valid queries; \textbf{Clause}, \textbf{Dep.}, and \textbf{Size}: the average of clauses, data dependencies, and bytes over all queries, respectively}
	\label{tbl_query_gen}
	\noindent{\small
		\begin{center}
			\begin{tabular}{@{}lrrrrr@{}}
			\toprule
\textbf{GDBMS} & \textbf{Gen.} & \textbf{Valid} & \textbf{Clause} & \textbf{Dep.} & \textbf{Size}\\
\midrule
Neo4j & 318k & 89.90\% & 40.97 & 26.77 & 1394.69\\
RedisGraph & 1'078k & 93.90\% & 32.71 & 30.59 & 1071.58\\
Memgraph & 570k & 83.27\% & 29.95 & 21.71 & 993.63\\
\cmidrule{1-6}
\textbf{Average} & 655k & 89.02\% & 34.54 & 26.36 & 1153.3\\
\bottomrule
\end{tabular}
\end{center}
}
\end{table}

\vspace{\midspace}
\noindent
\textbf{Query Validity}. 
Among the 655k queries generated by \sys, 583k are valid. The percentage of valid queries is 89.02\%. The result demonstrates that \sys can keep a high validity rate even when generating complex queries. We investigated the invalid queries generated by \sys and found that they were mainly caused by illegal arithmetic operations.
Some arithmetic operations require the operands to satisfy some constraints (\emph{e.g.}, the divisor in a division must be non-zero). However, \sys cannot track the value of each operator in queries because the value may depend on complex expressions, whose results are difficult to calculate (\emph{e.g.}, hash functions). Queries failing to satisfy the constraints of operations cause semantic errors, such as divisions by zero.

\vspace{\midspace}
\noindent
\textbf{Throughput}. On average, one \sys instance generates 3.79 tests (each test consists of one original query and one transformed query) per second. Specifically, at each second, one \sys instance completes 1.84 tests for Neo4j, 6.24 for RedisGraph, and 3.30 for Memgraph. 
To understand the bottleneck of test throughput, 
we further investigated the time used by Neo4j in query generation, transformation, and execution. We found that for Neo4j, 74.75\% of CPU time was used to execute the generated queries, which are complex and can be time-consuming.
This result indicates that the test throughput is dominated by the performance of the tested GDBMSs.
We believe the current throughput is practical considering (1) GDBMS testing typically lasts for several months~\cite{matteo2023issta, jiang2023icse}, and thus a sufficient number of test cases can be executed; and (2) setting up multiple \sys instances can significantly improve the test efficiency.

\subsection{Importance of Key Techniques}
\label{sec_sensitivity_analysis}

To show the importance of our techniques, we analyze how they contribute to the 127 bugs found by \sys.
The analysis consists of two parts.
The first focuses on the query context and graph schema in query generation, while the second assesses the transformation strategies (4 clause-level ones and 2 expression-level ones) used in test-oracle construction. 

We analyzed the \bugsfound bug-triggering queries and checked whether their generations depend on specific graph states. Specifically, we extracted the data dependencies contained by each query and check whether at least one dependency is related to query context or graph schema. Table~\ref{tbl_bug_graph_state} shows the results.
Among the \bugsfound queries, 72 depend on only query context, 5 depends on only graph schema, and 9 depend on both.
According to these results, without query context, 81 bugs (12 logic bugs) cannot be found. This demonstrates the importance of query contexts, which enable \sys to generate queries referencing specific nodes, relationships, or expressions constructed in internal clauses.
Without graph schema, 14 bugs (3 logic bugs) will be missed. These bug-triggering queries require \sys to properly reference the latest graph labels or properties.

\begin{table}[t]
	\caption{Bug-triggering queries related to \textbf{QC}: only query context; \textbf{GS}: only graph schema; \textbf{Both}: both graph states. The number in the parentheses is for the logic bugs
    }
	\label{tbl_bug_graph_state}
	\noindent{\small
		\begin{center}
			\begin{tabular}{@{}lrrrr@{}}
			\toprule
\textbf{Version} & \textbf{Found} & \textbf{QC} & \textbf{GS} & \textbf{Both}\\
\midrule
Neo4j & 51 (13) & 23 (3) & 3 (0) & 5 (2)\\
RedisGraph & 62 (18) & 34 (7) & 1 (1) & 4 (0)\\
Memgraph & 14 (2) & 5 (0) & 1 (0) & 0 (0)\\
\cmidrule{1-5}
\textbf{Total} & 127 (33) & 72 (10) & 5 (1) & 9 (2)\\
\bottomrule
\end{tabular}
\end{center}}
\end{table}

\begin{table}[t]
	\caption{Transformation strategies used in finding logic bugs, including Dead Code (DC), Unused Context (UC), Redundant Write (RW), Inconsequential Supersession (IS), Mathematical Identity (MI), and Cypher Feature (CF)}
	\label{tbl_logic_bug_strategies}
        \setlength\tabcolsep{3.3pt}
	\noindent{\small
		\begin{center}
			\begin{tabular}{@{}lcrrr|r@{}}
			\toprule
\textbf{Type} & \textbf{Strategy} & \textbf{Neo4j} & \textbf{RedisGraph} & \textbf{Memgraph} & \textbf{Total}\\
\midrule
\multirow{5}{*}{Clause} & DC & 3 & 4 & 0 & 7\\
& UC & 2 & 3 & 0 & 5\\
& RW & 1 & 3 & 1 & 5\\
& IS & 2 & 6 & 0 & 8\\
\cmidrule{2-6}
& Total & 8 & 16 & 1 & 25\\
\midrule
\multirow{3}{*}{Expr} & MI & 1 & 1 & 1 & 3\\
& CF & 4 & 1 & 0 & 5\\
\cmidrule{2-6}
& Total & 5 & 2 & 1 & 8\\
\midrule
\textbf{Total} & & 13 & 18 & 2 & 33\\
\bottomrule
\end{tabular}
\end{center}}
\end{table}

We investigated the 33 logic bugs found by \sys, and checked whether their bug-triggering queries use specific transformation strategies. 
Table~\ref{tbl_logic_bug_strategies} shows the results.
We find that each logic bug was triggered by using just one transformation rule in Table~\ref{tbl_clause_transformations} and Table~\ref{tbl_expression_transformations}. Specifically, among the 33 logic bugs, 25 are exposed by clause-level transformations, and 8 by expression-level transformations. The number of bugs found by each transformation strategy is close to the average (\emph{i.e.}, 5.5). Inconsequential Supersession finds the most bugs with 8, and Mathematical Identity finds the least with 3 bugs. These results indicate that all strategies are effective in identifying logic bugs in GDBMSs. Moreover, the 25 bugs found by clause-level transformations demonstrate the effectiveness of our fine-grained transformation over EET~\cite{zuming2024osdi}, which supports only expression-level transformation.

The following discusses two representative bug examples.

\vspace{\midspace}
\noindent
\textit{Example 1:} 
Figure~\ref{fig_neo4j_bug_RW_dependencies} shows a Neo4j bug that causes an incorrect query result. The original query references both query contexts and graph schema. Specifically, the first \cypher{CREATE} clause creates two nodes with a relationship \cypher{y}, labeled \cypher{A}. One node has a property \cypher{x} with value \cypher{0}. The second \cypher{CREATE} clause creates a new node with a property \cypher{x} whose value is equal to the property \cypher{x} of the end node of \cypher{y} (\emph{i.e.}, \cypher{0}). The \cypher{RETURN} clause returns the number of nodes whose property \cypher{x} is \cypher{0}. The expected result should be 2. However, Neo4j produces 1, because of a logic bug triggered when Neo4j processes two consecutive \cypher{CREATE}s that have data dependencies. \sys finds this bug using a clause-level transformation, Redundant Write. It inserts a \cypher{CREATE} clause and \cypher{DELETE} clause between the two \cypher{CREATE}s. Such a transformation preserves the query semantics but makes the two \cypher{CREATE}s not consecutive anymore, and thus Neo4j produces a correct result for the transformed query. The developers fixed this bug by significantly reconstructing the source code related to \cypher{CREATE} clauses. In the end, 17 source files were modified. 

\begin{figure}[t]
\noindent
\begin{minipage}[t]{0.47\linewidth}
    \begin{lstlisting}[
    language=cypher, 
    basicstyle=\footnotesize\tt,
]                              
// (|\textbf{Original: {1}}|) (|\bugmark|)
CREATE ({x:0})<-[y:A]-() 
(|\hl{\highlightedCREATE{} (\{x:endNode(y).x\})}|)
RETURN COUNT {({x:0})} AS n
\end{lstlisting}
\end{minipage}
\hfill
\begin{minipage}[t]{0.47\linewidth}
    \begin{lstlisting}[
    language=cypher, 
    basicstyle=\footnotesize\tt,
]        
// (|\textbf{Transformed: {2}}|) (|\textcolor{mygreen}{\tikzcheckmark}|)
CREATE ({x:0})<-[y:A]-() 
(|\hl{\highlightedCREATE{} (z) \highlightedDELETE{} z}|)
(|\hl{\highlightedCREATE{} (\{x:endNode(y).x\})}|)
RETURN COUNT {({x:0})} AS n
\end{lstlisting}
\end{minipage}
    \caption{Clause-level transformation reveals a bug.}
    \label{fig_neo4j_bug_RW_dependencies}
\end{figure}

\vspace{\midspace}
\noindent
\textit{Example 2:}
Figure~\ref{fig_neo4j_bug_no_dependencies} shows a logic bug in Neo4j found by expression-level transformation, Mathematical Identity. It is based on a rule that the result of \cypher{NULL XOR a} is \cypher{NULL}, no matter what \cypher{a} is. \sys transforms the \cypher{NULL} into \cypher|NULL XOR EXISTS{()}|, expecting that the result should remain \cypher{NULL}. However, Neo4j produces \cypher{FALSE}. The root cause of this bug is related to an operator in the Cypher planner of Neo4j, LetSelectOrSemiApply. It is invoked only when the transformed query is processed. Such an operator fails to work correctly because it does not consider \cypher{NULL} expressions in its arguments. To fix this bug, Neo4j developers refined this operator by adding additional logic to check and handle \cypher{NULL} in its arguments.


\begin{figure}[t]
\noindent
\begin{minipage}[t]{0.42\linewidth}
    \begin{lstlisting}[
    language=cypher, 
    basicstyle=\footnotesize\tt,
]                              
// (|\textbf{Original: {NULL}}|) (|\textcolor{mygreen}{\tikzcheckmark}|)
RETURN (|\hl{NULL}|) AS x
\end{lstlisting}
\end{minipage}
\hfill
\begin{minipage}[t]{0.52\linewidth}
    \begin{lstlisting}[
    language=cypher, 
    basicstyle=\footnotesize\tt,
]        
// (|\textbf{Transformed: {FALSE}}|) (|\bugmark|)
RETURN (|\hl{NULL XOR \highlightedEXISTS \{()\}}|) AS x
\end{lstlisting}
\end{minipage}
    \caption{Expression-level transformation reveals a bug.}
    \label{fig_neo4j_bug_no_dependencies}
\end{figure}

\subsection{Comparison}
\label{sec_comparison}

\textbf{Bug Latency Study.}
To demonstrate that \sys can find bugs missed by existing approaches, 
we investigate the latencies of bugs found by \sys.
For each bug, we check whether its bug-inducing commit was created before the years when existing approaches were published. 
If \sys finds some long-latent bugs, we can conclude that \sys can find bugs missed by existing approaches.
This comparison is reasonable and objective because:
(1) none of the bugs found by \sys are marked as duplicated by developers, meaning that no approach found these bugs until \sys found them;
and (2) all existing approaches have extensively tested Neo4j and RedisGraph~\cite{matteo2023issta, ziyue2023issta, jiang2023icse, zhuang2023vldb, qiuyang2024grev}, meaning that in these two GDBMSs, no approach found the long-latent bugs found by \sys during their evaluation. Some existing approaches do not test Memgraph, but we still include it to show the effectiveness of \sys.

GDsmith~\cite{ziyue2023issta}, GDBMeter~\cite{matteo2023issta}, GraphGenie~\cite{jiang2023icse}, and GAMERA~\cite{zhuang2023vldb} were published in 2023, and GRev~\cite{qiuyang2024grev} was published in 2024. Therefore, for each bug, we checked whether its bug-inducing commit was created before 2024 (\emph{i.e.}, in 2023 or earlier) and 2023. As shown in Table~\ref{tbl_bug_latency}, among the 113 bugs in Neo4j and RedisGraph, 47 already existed before 2023, among which 10 are logic bugs. It indicates that all existing approaches failed to find these 47 bugs in their extensive evaluation. 
In addition, 45 bugs (24 logic bugs) were introduced to Neo4j and RedisGraph between 2023 and 2024, while GRev, which was proposed after 2024, cannot find these bugs. In addition, GDsmith and GRev support Memgraph, but neither can find the 4 long-latent bugs induced before 2023.
These results indicate that existing approaches indeed miss many long-latent bugs, while \sys can effectively find them.


\begin{table}[t]
	\caption{Latency of the bugs found by \sys. The number in the parentheses is for the logic bugs}
	\label{tbl_bug_latency}
	\noindent{\small
		\begin{center}
			\begin{tabular}{@{}lrrrr@{}}
			\toprule
\multirow{2}{*}[-0.7ex]{\textbf{DBMS}} & \multirow{2}{*}[-0.7ex]{\textbf{Found}} & \multicolumn{2}{c}{\textbf{Bug-involved year}}  & \multirow{2}{*}[-0.7ex]{\textbf{Longest latency}}\\
\cmidrule(l){3-4}
& & \textbf{< 2024} & \textbf{< 2023} & \\
\midrule
Neo4j & 51 (13) & 40 (9) & 27 (7) & 2016 \\
RedisGraph & 62 (18) & 52 (15) & 20 (3)  & 2020 \\
Memgraph & 14 (2) & 9 (2) & 4 (0) & 2021\\
\cmidrule{1-5}
\textbf{Total} & 127 (33) & 101 (26) & 51 (10) & 2016\\
\bottomrule
\end{tabular}
\end{center}}
\end{table}

\begin{table}[t]
	\caption{The clauses used by bug-triggering queries. The number in the parentheses is for the logic bugs. N/A indicates that the GDBMS does not support the clause}
	\label{tbl_clause_bugs}
	\noindent{\small
		\begin{center}
  \setlength{\tabcolsep}{3.1pt}
			\begin{tabular}{lrrr|r}
			\toprule
\textbf{Clause} & \textbf{Neo4j} & \textbf{RedisGraph} & \textbf{Memgraph} & \textbf{Total} \\
\midrule
\cypher|MATCH|  & 20 (7) & 28 (6) & 5 (1) & 53 (14) \\
\cypher|CREATE| & 17 (9) & 26 (11) & 2 (1) & 55 (21) \\
\cypher|MERGE|   & 12 (4) & 31 (12) & 5 (1) & 48 (17) \\
\cypher|DELETE|  & 4 (4) & 16 (6) & 1 (1) & 21 (11) \\
\cypher|REMOVE|  & 0 & 1 (0) & 0 & 1 (0)\\
\cypher|SET|     & 4 (2) & 3 (1) & 0 & 7 (3) \\
\cypher|UNWIND| & 9 (1) & 11 (6) & 1 (0) & 21 (7) \\
\cypher|WITH|   & 16 (3) & 23 (8) & 0 & 39 (11) \\
\cypher|RETURN| & 40 (8) & 39 (13) & 11 (1) & 90 (22) \\
\cypher|CALL|    & 15 (3) & 22 (8) & 4 (0) & 41 (11) \\
\cypher|FOREACH| & 8 (5) & 6 (3) & 0 & 14 (8) \\
\cypher|UNION|   & 10 (3) & 7 (3) & 1 (0) & 18 (6) \\
\cypher|EXISTS|  & 10 (3) & N/A & N/A & 10 (3) \\
\cypher|COUNT|   & 10 (4) & N/A & N/A & 10 (4) \\
\bottomrule
\end{tabular}
\end{center}}
\end{table}

\vspace{\midspace}
\noindent
\textbf{Clause Analysis.}
We analyze the clauses used in the 127 bug-triggering queries and show the results in Table~\ref{tbl_clause_bugs}. The results indicate that the bugs found by \sys are related to various Cypher clauses. Combining Table~\ref{tbl_supported_clause} and Table~\ref{tbl_clause_bugs}, we can demonstrate that many bugs found by \sys cannot be found by existing approaches, because they do not even support the clauses used in some bug-triggering queries (\emph{e.g.}, 14 bugs (8 logic bugs) related to \cypher{FOREACH} clauses).
As discussed in Section~\ref{sec_implementation}, existing approaches cannot support these clauses because they lack a systematic model for query generation and a general test oracle for query validation.

\vspace{\midspace}
\noindent
\textbf{Empirical Comparison.} To empirically demonstrate that \sys outperforms the state-of-the-art on code coverage and bug detection,
we further evaluated \sys and existing approaches on Neo4j and RedisGraph, which are the only two GDBMSs supported by all these approaches. Each evaluation persisted for 48 hours and was repeated 5 times.
Table~\ref{tbl_comparison} shows the comparison results.

\begin{table}[t]
	\caption{Results of covered lines (average $\pm$ standard deviation) and found bugs of existing approaches over 5 runs }
	\label{tbl_comparison}
        \setlength\tabcolsep{4pt}
	\noindent{\small
		\begin{center}
			\begin{tabular}{lrrrr}
			\toprule
\multirow{2}{*}[-0.7ex]{\textbf{}} & \multicolumn{2}{c}{\textbf{Neo4j}} & \multicolumn{2}{c}{\textbf{RedisGraph}} \\
\cmidrule(l){2-3}  \cmidrule(l){4-5}
 & \textbf{Line (Avg $\pm$ SD)} & \textbf{Bug} & \textbf{Line (Avg $\pm$ SD)} & \textbf{Bug} \\
\midrule
\sys & 750k $\pm$ 5.2k & 19 & 20k $\pm$ 0.1k & 26 \\
GDBMeter & 283k $\pm$ 0.2k & 1 & 19k $\pm$ 0.3k & 1 \\
GRev & 597k $\pm$ 4.0k & 1 & 14k $\pm$ 0.1k & 1 \\
GAMERA & 283k $\pm$ 0.1k & 1 & 3k $\pm$ 0.1k & 0 \\
GDsmith & 515k $\pm$ 8.6k & 0 & 15k $\pm$ 0.1k & 0 \\
GraphGenie & 506k $\pm$ 2.0k & 0 & 3k $\pm$ 0.1k & 0 \\
\bottomrule
\end{tabular}
\end{center}}
\end{table}

\vspace{\midspace}
\textit{Code Coverage.} As shown in Table~\ref{tbl_comparison}, on average, \sys covers 71\% and 85\% more code than existing approaches in Neo4j and RedisGraph, respectively.
Among the tools, excluding \sys, GRev covers the most code (\emph{i.e.}, 597k lines) in Neo4j. Compared to GRev, \sys covers 25\% more code. Benefiting from the powerful query generation, \sys can cover much deeper logic of GDBMSs that is related to processing advanced Cypher features and complicated data dependencies. In contrast, simple queries generated by existing tools rarely trigger such logic.
In RedisGraph, excluding \sys, GDBMeter covers the most code (\emph{i.e.}, 19k lines). Compared to it, \sys covers 5\% more code. The coverage improvement is less significant than in Neo4j, because RedisGraph supports fewer Cypher features and cannot handle some complicated Cypher semantics. For example, RedisGraph does not support processing the subqueries in \cypher|FOREACH| and \cypher|CALL| clauses. As a result, the space for \sys to improve the code coverage in RedisGraph is limited. 

\begin{figure}[t]
	\centering\includegraphics[width=0.75\linewidth]{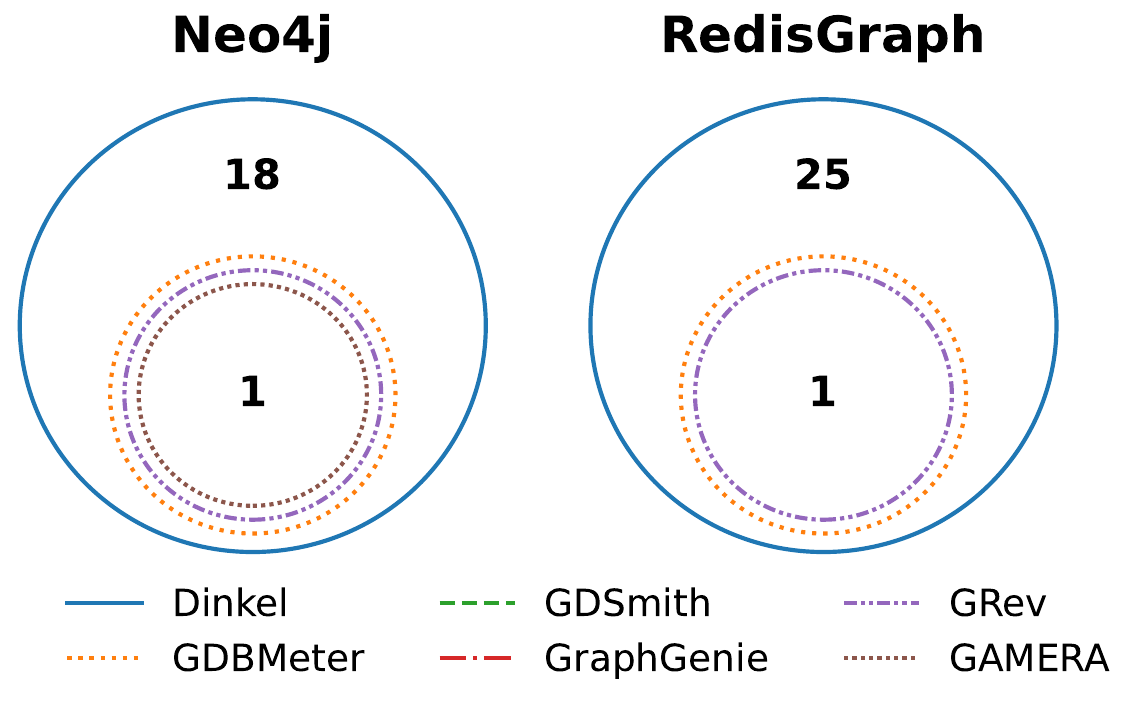}
	\caption{Relation of bugs found by existing approaches.} 
	\label{fig_bug_comparison}
\end{figure}

\vspace{\midspace}
\textit{Found Bugs.} 
Table~\ref{tbl_comparison} shows the number of bugs found by each approach over 5 runs.
Figure~\ref{fig_bug_comparison} describes the bug relation. 
\sys finds the most bugs in both Neo4j (19 bugs) and RedisGraph (26 bugs), where logic bugs comprise 4 of the bugs found in Neo4j and 7 in RedisGraph.
In Neo4j, GDBMeter, GRev, and GAMERA found 1 bug.
In RedisGraph, GRev and GDBMeter found 1 bug. Both of these bugs were also identified by \sys. 
For the 43 bugs missed by existing approaches, their bug-triggering queries contain either complicated data dependencies or advanced Cypher features that are not supported by existing approaches. 
The bugs shown in Figure~\ref{fig_neo4j_bug_RW_dependencies} and Figure~\ref{fig_neo4j_bug_no_dependencies} are two of the 43 bugs. 
These results demonstrate that \sys can find more bugs than existing approaches by generating and validating complex queries.

\section{Related Work}


\noindent\textbf{GDBMS Testing.} 
GDBMS testing is an emerging research field, where several approaches~\cite{jiang2023icse, matteo2023issta, ziyue2023issta, yingying2022issta, zhuang2023vldb, qiuyang2024grev} have been proposed. All these approaches aim to
construct test oracles to identify logic bugs. Grand~\cite{yingying2022issta} targets GDBMSs using the Gremlin query language and detects logic bugs using differential testing~\cite{mckeeman1998differential}. 
Several approaches~\cite{jiang2023icse, matteo2023issta, ziyue2023issta, zhuang2023vldb, qiuyang2024grev} support GDBMSs using Cypher.
Similar to Grand, GDsmith~\cite{ziyue2023issta} also uses differential testing to find logic bugs. GDBMeter~\cite{matteo2023issta} leverages query partitioning~\cite{manuel2020oopsla}, which decomposes the predicate of a query and checks if the queries with the decomposed predicates produce consistent results with the original query. GraphGenie~\cite{jiang2023icse} modifies the graph patterns used in a query and checks if the query with modified graph patterns satisfies the expected relationship (\emph{i.e.}, equivalence, subset, or superset) with the original queries. GAMERA~\cite{zhuang2023vldb} extracts metamorphic relations~\cite{chen2020metamorphic} from the manipulated graph data and checks whether the outputs of generated queries satisfy these relations.

\sys is different from all existing approaches in two key aspects. First, 
\sys significantly improves query generation, the core foundation of GDBMS testing. 
Second, while existing approaches support correctness testing for only queries following specific patterns, \sys generally tackles the problems of test-oracle construction, enabling validation for arbitrary queries.


\vspace{\midspace}
\noindent\textbf{RDBMS Testing.}
Compared to GDBMS testing, testing relational database management systems (RDBMSs) is more mature, where both query generation~\cite{sqlsmith, rui2020ccs, fu2022griffin, zuming2023sec, liang2023sequence} and test-oracle construction~\cite{manuel2020osdi, manuel2020fse, manuel2020oopsla, zongyin2023atc, jiansen2023icse, xiu2023sigmod, zuming2024osdi} are well-researched.
SQLsmith~\cite{sqlsmith} embeds the SQL grammar~\cite{sql_standard} and can generate complex SQL statements. 
DynSQL~\cite{zuming2023sec} incrementally generates SQL statements for a query by querying the latest DBMS schema, which enables DynSQL to decouple the generation of each statement and helps it generate queries containing multiple complex statements. 
To find logic bugs in RDBMSs, PQS~\cite{manuel2020osdi} synthesizes customized queries that fetch specific rows of tables. If the tested RDBMS fails to fetch the rows, a logic bug is identified. Pinolo~\cite{zongyin2023atc} modifies the predicate of a query, where the modified query should produce the subset/superset of the results of the original query. Otherwise, Pinolo identifies a logic bug.
EET~\cite{zuming2024osdi} transforms the expressions of queries in a semantic-preserving manner and checks whether the queries with the transformed expressions produce the same results as the original queries. EET identifies a bug if their results are not the same. Due to the separation of querying and data manipulation in the SQL language, EET cannot perform clause-level transformations the same way \sys can for Cypher queries.

Unlike these approaches for SQL queries, \sys is designed for validating Cypher queries. Leveraging the features of the Cypher query language, \sys integrates novel techniques for both query generation and test-oracle construction.


\vspace{\midspace}
\noindent\textbf{State-Aware Fuzzing.}
Fuzzing is a promising technique for finding bugs in software~\cite{zuming2020sec, zuming2022ndss, AFL, chenyang2019sec, Yun2018sec, chen2018angora}.  
Some approaches~\cite{ma2023ndss, kim2022sp, zuming2023sec, zhao2022sec, atlidakis2019icse} have been proposed to find bugs more efficiently in state-sensitive systems. RESTler~\cite{atlidakis2019icse} analyzes API specifications of the tested cloud services and generates request sequences that follow the inferred producer-consumer dependencies. RESTler also collects the response observed during prior requests to guide subsequent request generation. 
LOKI~\cite{ma2023ndss} proposed to test blockchain consensus protocols. It builds a state model to dynamically track the state transition of each node in the blockchain systems and accordingly generates inputs with proper targets, types, and contents.
StateFuzz~\cite{zhao2022sec} is designed to test Linux drivers. It
utilizes static analysis to recognize critical variables that affect control flows or memory accesses, and represents program states using these variables. StateFuzz prioritizes test cases that trigger new states. For effectively testing USB gadget stacks, FuzzUSB~\cite{kim2022sp} extracts the internal state machines from USB gadget drivers via static analysis and symbolic execution, before using this state information as fuzzing feedback for guiding test-case generation.

Different from these approaches, \sys models GDBMS-related state information for allowing query generation to introduce complex data dependencies, all while retaining query validity.
\section{Conclusion}

We have presented a novel and practical framework, \sys, for validating GDBMSs. We model the graph state as query context and graph schema, and propose state-aware query generation to generate complex and valid Cypher queries for testing the deep logic of GDBMSs. Moreover, we propose two fine-grained query transformations: clause-level transformations and expression-level transformations, which can operate on arbitrary Cypher queries to validate their correctness.
In our evaluation, \sys found \bugsfound bugs in three well-known GDBMSs, among which \bugslogic are logic bugs.
Considering its significant advancement in bug finding, we believe \sys can lay a practical foundation for GDBMS testing, facilitating and inspiring follow-up research on GDBMS reliability.

\bibliographystyle{ACM-Reference-Format}
\balance
\bibliography{sample-base}

\end{document}